\documentclass[10pt, twocolumn, letterpaper]{article}
\usepackage{usenix}

\usepackage{tikz}
\usetikzlibrary{calc}
\usepackage{amsmath,amssymb,amsfonts}
\usepackage{xcolor}
\usepackage[noend]{algpseudocode} \usepackage{algorithm}
\newfloat{algorithm}{t}{lop}
\usepackage{float}
\usepackage{makecell}
\usepackage{multirow}
\usepackage{subcaption}
\usepackage{listings}
\usepackage{relsize}
\usepackage{url}

\usepackage{booktabs}
\usepackage{csquotes}
\usepackage{cleveref}
\usepackage{enumitem}
\usepackage{adjustbox}

\graphicspath{ {./figs/} }

\usepackage[scale=0.95]{inconsolata}

\definecolor{bluekeywords}{rgb}{0.13,0.13,1}
\definecolor{greencomments}{rgb}{0.25,0.5,0.35}
\definecolor{deepred}{rgb}{0.6,0,0}
\definecolor{deepblue}{rgb}{0,0,0.5}
\definecolor{deepgreen}{rgb}{0,0.5,0}

\hypersetup{
    colorlinks, linkcolor={deepred},
    citecolor={deepgreen},
    urlcolor={deepblue}
}

\lstdefinelanguage{WebAssembly}{
  sensitive=true,
  otherkeywords={},
  morekeywords=[1]{i32,f32,i64,f64},
  keywordstyle={[1]\color{violet}},
  morekeywords=[2]{0},
  keywordstyle={[2]\color{violet}},
  morekeywords=[3]{add,const}
  keywordstyle={[3]\color{bluemunsell}},
  morekeywords=[4]{},
  keywordstyle={[4]\color{candypink}},
  morekeywords=[5]{module, func, param, result, global, mut, export, import, memory, data, local, elem, table, call, call_indirect, type},
  keywordstyle={[5]\color{bluekeywords}},
  morekeywords=[6]{=,;},
  keywordstyle={[6]\color{britishracinggreen}},
  morekeywords=[7]{(,),[,],.},
  keywordstyle={[7]\color{black}},
  rulecolor=\color{black},
  morecomment=**[l][\color{greencomments}]{;;},
  stepnumber=1,
  escapechar=|,
}

\definecolor{halfgray}{gray}{0.5}
\definecolor{lightgray}{gray}{0.7}

\newcolumntype{?}{!{\vrule width 2pt}}
\definecolor{DARKRED}{rgb}{0.75,0,0}

\newcommand{\parag}[1]{\smallskip\noindent{}\textbf{#1.}\quad}

\newcommand{\name}{\textsf{Fuzzm}} \newcommand{\afl}{AFL}

\newcommand{\code}[1]{\text{\lstinline[basicstyle=\ttfamily]~#1~}}

\usepackage{tcolorbox}
\newenvironment{result}{
\begin{tcolorbox}[colback=blue!5!white,colframe=blue!5!white,arc=0mm,grow to left by=1.5mm,left=0mm,grow to right by=1.5mm,right=0mm,top=0mm,bottom=0mm]
\textbf{Summary:}
}
{
\end{tcolorbox}
}

\usepackage{titlesec}

\titlespacing\section{0pt}{10pt plus 4pt minus 2pt}{8pt plus 2pt minus 2pt}
\titlespacing\subsection{0pt}{8pt plus 4pt minus 2pt}{6pt plus 2pt minus 2pt}
\titlespacing\subsubsection{0pt}{6pt plus 4pt minus 2pt}{3pt plus 2pt minus 2pt}

\newcolumntype{C}{@{\extracolsep{1pt}}c@{\extracolsep{1pt}}}

\begin{document}

\date{}

\title{\name{}: Finding Memory Bugs through \\Binary-Only Instrumentation and Fuzzing of WebAssembly}

\author{
{\rm Daniel Lehmann}\thanks{\protect\refstepcounter{footnote}\protect\label{thanks}Both authors contributed equally to the paper.}\addtocounter{footnote}{-1}\addtocounter{Hfootnote}{-1}\\
  University of Stuttgart,\\Germany\\
  \texttt{mail@dlehmann.eu}
  \and
  {\rm Martin Toldam Torp}\textsuperscript{\ref{thanks}}\\
  Aarhus University,\\Denmark\\
  \texttt{torp@cs.au.dk}
  \and
  {\rm Michael Pradel}\\
  University of Stuttgart,\\Germany\\
  \texttt{michael@binaervarianz.de}
} 

\maketitle

\begin{abstract}
WebAssembly binaries are often compiled from memory-unsafe languages, such as C and C++.
Because of WebAssembly's linear memory and missing protection features, e.g., stack canaries, source-level memory vulnerabilities are exploitable in compiled WebAssembly binaries, sometimes even more easily than in native code.
This paper addresses the problem of detecting such vulnerabilities through the first binary-only fuzzer for WebAssembly.
Our approach, called \name{}, combines canary instrumentation to detect overflows and underflows on the stack and the heap, an efficient coverage instrumentation, a WebAssembly VM, and the input generation algorithm of the popular AFL fuzzer.
Besides as an oracle for fuzzing, our canaries also serve as a stand-alone binary hardening technique to prevent the exploitation of vulnerable binaries in production.
We evaluate \name{} with 28 real-world WebAssembly binaries, some compiled from source and some found in the wild without source code.
The fuzzer explores thousands of execution paths, triggers dozens of crashes, and performs hundreds of program executions per second.
When used for binary hardening, the approach prevents previously published exploits against vulnerable WebAssembly binaries while imposing low runtime overhead.
\end{abstract}

\section{Introduction}
\label{sec:introduction}

WebAssembly is an increasingly important bytecode language~\cite{haas2017bringing, wasm-spec, watt2018mechanising} with low-level semantics, fast execution, and a multitude of source languages compiling to it.
It is widely supported in browsers\footnote{>\,94\% support as of October 2021, see \url{https://caniuse.com/wasm.}} and used by diverse web applications~\cite{wasm-empirical}, for \enquote{serverless} cloud computing~\cite{254432, 10.1145/3302505.3310084}, in smart contract platforms~\cite{ewasm-medium, quan2019evulhunter, he2020security}, to sandbox libraries in native applications~\cite{narayan2020retrofitting, wasm-box-c}, and even as a universal bytecode by standlone WebAssembly runtimes~\cite{wasi-mozilla-hacks, wasmer-website, wasmtime-website}.

Given its importance, the security of WebAssembly is also becoming more and more relevant.
While WebAssembly prevents some security issues by design, source-level vulnerabilities may still propagate to WebAssembly binaries~\cite{McFadden2018, forcepoint, DBLP:conf/uss/0002KP20}.
Recent work~\cite{DBLP:conf/uss/0002KP20} has shown that, surprisingly, memory vulnerabilities in WebAssembly binaries can sometimes be even more easily exploited than when the same source code is compiled to native architectures.
One reason is the lack of mitigations, such as stack canaries, page protection flags, or hardened memory allocators~\cite{DBLP:conf/uss/0002KP20}.

To find vulnerabilities, \emph{greybox fuzzing} has proven to be an effective technique~\cite{Boehme2019, DBLP:journals/corr/abs-2009-01120, rawat2017vuzzer, li2017steelix, zeller2019fuzzing}.
For example, Google's OSS-Fuzz project has found thousands of vulnerabilities in widely used software~\cite{serebryany2017oss, oss-fuzz}.
A greybox fuzzer automatically generates inputs that explore the target program and eventually trigger a vulnerability.
For that, it requires (i) lightweight feedback from the execution, e.g., coverage information, to guide the input generation, and (ii) runtime oracles that make a vulnerability apparent, e.g., by crashing the program.

A greybox fuzzer for WebAssembly would be highly desirable, but several characteristics of WebAssembly must be taken into account.
First, WebAssembly is a compilation target for multiple source languages, including C, C++, Rust, Go, and many others~\cite{wasm-empirical}.
A fuzzer aimed at a specific source language hence could analyze only a fraction of all real-world binaries.
Second, the source code of a WebAssembly binary may not be available, e.g., when analyzing third-party websites, third-party libraries, or in-house legacy applications.
Even if the source code is available, adopting a fuzzer into the development workflow is made harder if it requires changes to the build system, or specific (versions of) compilers.
Third, even when compiling from the same source code, the security-relevant behavior of a program compiled to WebAssembly may differ from the same program being compiled to native code~\cite{DBLP:conf/uss/0002KP20}.
As we illustrate in Section~\ref{sec:overview}, whether a vulnerability can be exploited depends on how the semantics of the source language are compiled and which mitigations the target platform provides.
As a result, fuzzing a program compiled for another platform, e.g., x86~\cite{Dinesh2020}, is insufficient to expose memory bugs in WebAssembly.
Taken together, these characteristics motivate a fuzzer targeted at WebAssembly binaries.
However, despite the overall success of greybox fuzzing and the increasing importance of WebAssembly, such a fuzzer currently does not exist.

\begin{figure}
  \includegraphics[width=\linewidth,trim=8mm 0mm 8mm 15mm]{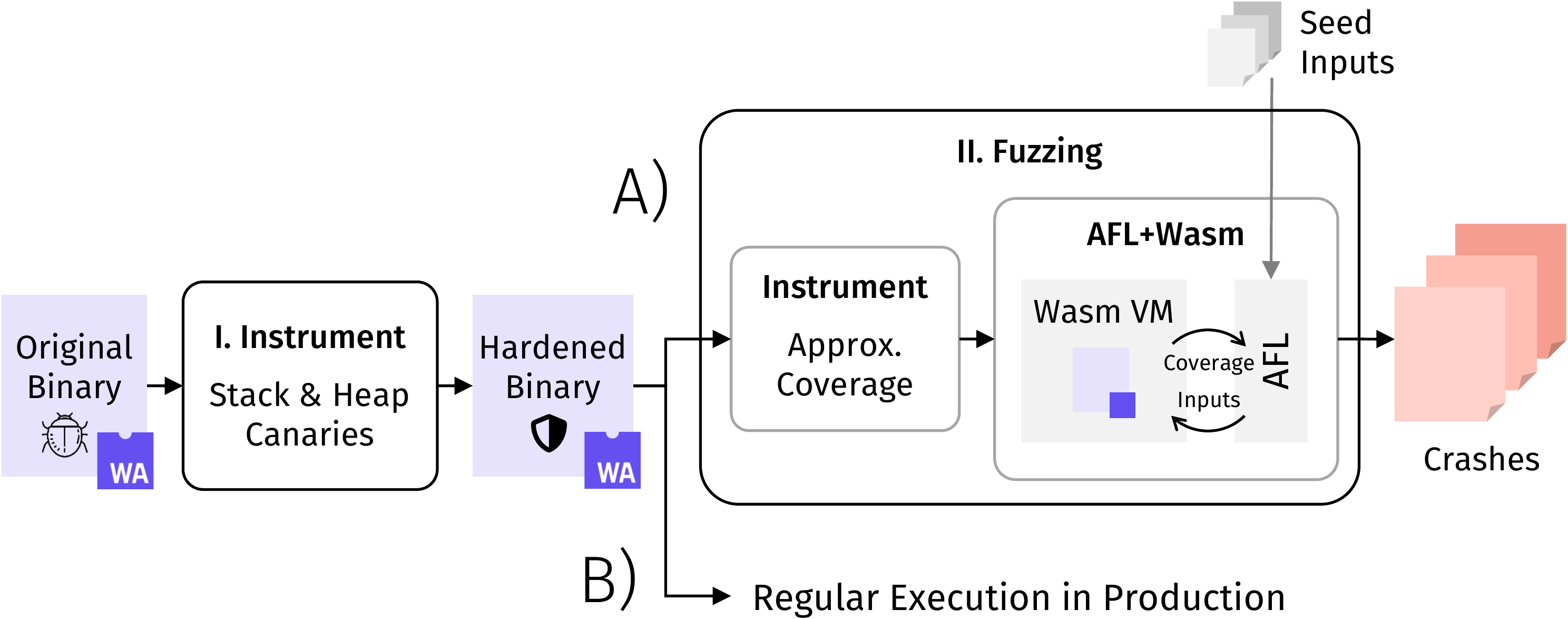}
  \caption{Overview of the main components of \name{}.}
  \label{fig:overview}
  \vspace{-2mm}
\end{figure}

This paper presents \name{}\footnote{``\name{}'' is a portmanteau word of ``fuzzing'' and ``Wasm''.}, the first binary-only greybox fuzzer for WebAssembly.
Its main components, shown in \Cref{fig:overview}, address several interesting technical challenges.
First, unlike native programs, WebAssembly lacks several built-in oracles that native fuzzers rely on for finding suspicious program behavior.
For example, none of the current compilers targeting WebAssembly adds stack canaries~\cite{DBLP:conf/usenix/PrasadC03, Cowan1998}, and due to WebAssembly's linear memory, overflows from, e.g., stack to heap data remain unnoticed~\cite{DBLP:conf/uss/0002KP20}.
While tools like AddressSanitizer~\cite{DBLP:conf/usenix/SerebryanyBPV12} can instrument source code to detect memory-related misbehavior, they do not apply to binaries.
Instead, our stack and heap canary instrumentation rewrites binaries to detect over- and underflows on the stack and heap.
Besides fuzzing, the canaries are also useful for retroactively hardening existing WebAssembly binaries in production.

Second, in a binary fuzzer we cannot rely on compiler-inserted code to track coverage, which is what AFL and other fuzzers do~\cite{Boehme2019, rawat2017vuzzer, li2017steelix}.
Even though there are dynamic instrumentation approaches for binaries, e.g., \afl{}'s QEMU mode, they often suffer from high overheads, and are architecture-dependent and not applicable to WebAssembly.
Our coverage instrumentation instead applies to unmodified, production WebAssembly binaries and tracks coverage efficiently.

The final challenge, especially when fuzzing bytecode programs, is efficiency.
WebAssembly binaries are executed in a virtual machine (VM), which may cause a naive approach to suffer from high start-up time and makes fuzzing impractical.
Instead, we integrate a WebAssembly VM that executes the target program with the tried-and-tested input generation of AFL.
Here, WebAssembly's sandboxing can actually be an opportunity rather than a drawback:
The memory of the target application and AFL can reside in a single address space, without the need for different processes separating the two.

The result of addressing the above challenges is a practical, effective, and efficient end-to-end fuzzer for WebAssembly binaries.
Our evaluation applies \name{} to 28 programs, of which ten are well-known programs compiled from source code to WebAssembly, and 18 are WebAssembly binaries without source code found in the wild.
We find our approach to be effective, covering 1,232 unique execution paths and triggering 40 unique crashes on average during 24 hours of fuzzing.
The majority of the triggered crashes are due to our canary-based oracles.
In terms of efficiency, \name{} performs hundreds of program executions per second, comparable to AFL, despite requiring only a binary as input and running the program in a VM.
Finally, we show that the canaries inserted by our instrumentations effectively prevent three previously published exploits against vulnerable WebAssembly binaries~\cite{DBLP:conf/uss/0002KP20}. 
Due to their low runtime overhead (1.05x and 1.06x, for stack and heap canaries, respectively) the canary instrumentation serves, beyond fuzzing, as a standalone hardening tool for existing, vulnerable WebAssembly binaries.

\paragraph{Contributions.}
In summary, this paper contributes:
\begin{itemize}[itemsep=-2pt,topsep=3pt,leftmargin=\parindent]
  \item The first \emph{binary-only fuzzer for WebAssembly} programs.
  \item A binary instrumentation that inserts stack and heap canaries, which can be used to \emph{harden existing WebAssembly programs} and as an \emph{oracle in our fuzzer} (\Cref{sec:canaries}).
  \item Integration of the AFL fuzzer and its tried-and-tested input generation, a binary-only instrumentation that provides compatible coverage information, and a WebAssembly VM, for \emph{efficient end-to-end fuzzing} (\Cref{sec:afl}).
  \item Empirical evidence that \name{} \emph{effectively explores paths and finds crashes} in well-known programs compiled to WebAssembly, and in large, real-world WebAssembly binaries without source code (\Cref{sec:evaluation}), and
  \item Empirical evidence that binaries hardened with our canary instrumentations run with \emph{low runtime overhead} and \emph{effectively thwart previously published exploits} (\Cref{sec:evaluation}).
\end{itemize}

\section{Background on WebAssembly}
\label{sec:background}

We briefly introduce features of WebAssembly most relevant to this paper, and some of the security aspects of the language.
WebAssembly is an assembly-like language designed as a portable compilation target from different source languages, e.g., from C/C++ with Emscripten or Clang, or from Rust.
Beyond the browser platform, which was the first to widely support WebAssembly, there now are several other platforms, e.g., Node.js and the standalone Wasmtime VM\footnote{\url{https://wasmtime.dev/}}.

\parag{Types}
WebAssembly is a stack-based, statically typed language.
There are four types in WebAssembly: \code{i32} (\code{i64}) for 32-bit (64-bit) integers and \code{f32} (\code{f64}) for 32-bit (64-bit) floating point numbers.
Typed instructions push and pop values from an implicit \emph{operand stack}.
For example, \code{i32.const} $N$ pushes the 32-bit constant integer $N$, and \code{f64.add} pops two 64-bit floating point numbers and then pushes their sum.

\begin{figure}
  \vspace{-2mm}
  \hfill
  \newcommand{\tikzmark}[2]{\tikz[overlay,remember picture] \node[text=black,
        inner sep=2pt] (#1) {#2};}
  \newcommand{\DrawArrows}[3][]{\coordinate (Start Mid) at ($(#2) + (0em,2pt)$);
      \coordinate (End Mid)   at ($(#3) + (0em,2pt)$);
      \draw[-stealth, thick, #1] (Start Mid) to (End Mid);
  }
  \begin{minipage}{.5\linewidth}
  \begin{lstlisting}[escapeinside={(*@}{@*)}]
loop (*@\tikzmark{end}{}@*)
  ...
  br 0 (*@\tikzmark{start}{}@*)  ;; Restart loop.
  ...
end(*@
\begin{tikzpicture}[overlay,remember picture]
\DrawArrows[in=5,out=10]{start}{end}
\end{tikzpicture}
@*)
  \end{lstlisting}
  \end{minipage}
  \hfill
  \begin{minipage}{.45\linewidth}
  \begin{lstlisting}[escapeinside={(*@}{@*)},firstnumber=6]
block
  block
    br 1 (*@\tikzmark{start}{}@*)  ;; Jump out.
  end
end (*@\tikzmark{end}{}@*)(*@
\begin{tikzpicture}[overlay,remember picture]
\DrawArrows[in=5,out=320]{start}{end}
\end{tikzpicture}
@*)
  \end{lstlisting}
  \end{minipage}
  \vspace{-3mm}
  \caption{Branching in WebAssembly illustrated}
  \label{fig:branching}
  \vspace{-3mm}
\end{figure}
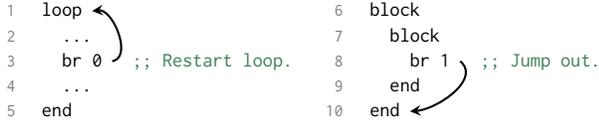

\parag{Control flow}
Unlike many other assembly-like languages, WebAssembly features structured control flow, which is encoded using nested blocks.
A block is a sequence of instructions that either begins with a \code{block} or \code{loop} and ends with an \code{end} instruction.\footnote{There are also \code{if} and \code{else} blocks, but they add no expressive power over regular blocks and branches, so we do not explain them here for brevity.}
Blocks may be nested arbitrarily deep.
Within a block, a \code{br} (\code{br\_if}) $L$ instruction (conditionally) jumps to the end (or for \code{loop}s, to the beginning) of the $L$th block, where 0 is the block containing the \code{br\_if} instruction, 1 is its parent, and so on.
$L$ can be thought of as a numerical, relative block \emph{label}.
For example, the \code{br} instruction on line~3 of \Cref{fig:branching} jumps to the beginning of the \code{loop} block on line~1, and similarly, the \code{br} on line~8 jumps out two blocks to the \code{end} on line~10.
A \code{br\_table} $L_0\dots{}L_t$, $D$ instruction implements jump tables.
It consumes the top-most integer from the operand stack $i$ and jumps to $L_i$, or a default $D$ if $i > t$.

\parag{Functions and variables}
A function in WebAssembly has typed parameters, typed local variables, and a sequence of instructions.
Parameters and local variables are read using \code{local.get} $N$ and written using \code{local.set} $N$, where $N$ refers to the $N$th local variable or parameter.
Most instructions, including direct calls, are statically type checked; indirect calls are type checked at runtime.
There are also global variables, which are read (written) with \code{global.get} (\code{global.set}).

\parag{Memory}
Unlike native programs, WebAssembly uses a byte-addressable \emph{linear memory} for storing long-lived objects.
The memory is initialized with a certain size when the module is instantiated and can be grown at runtime using the \code{memory.grow} instruction.
The 32-bit address space has no holes, so every pointer $\in$ [0, \emph{size}] is valid.
The memory is fully program-organized, i.e., there is no garbage collection.
Load and store instructions take \code{i32} values as addresses.
For example, \code{i64.store} consumes two elements from the operand stack: an \code{i64} value and an \code{i32} address, at which it stores the eight byte value in linear memory.

\parag{WASI}
WebAssembly does not have a standard library or I/O functions by default.
Instead all interaction with the underlying host system needs to happen through imports.
WASI (the WebAssembly System Interface) specifies a syscall interface of functions for performing I/O, filesystem access, etc.\footnote{\url{https://wasi.dev/}}
A program written in some high-level language (typically C, C++, or Rust) can be compiled to a WASI binary, and then executed by a WASI-compliant runtime, such as Wasmtime.
There are other ways of executing WebAssembly, e.g., in the browser or on Node.js, but we focus on WASI in this work.

\parag{Security}
WebAssembly has a two-sided security story.
On the one hand, WebAssembly programs execute in a sandboxed environment, which isolates them well from the memory and code of the underlying host system.
This prevents many attacks in the browser or cloud setting.
On the other hand, protection of WebAssembly program's \emph{own} memory is very limited~\cite{DBLP:conf/uss/0002KP20, chasm, forcepoint}, even compared to native binaries.
As the operand stack only stores primitive values, linear memory must be used for all non-primitive values on the stack, all static data, and the heap.
Current compilers do not insert canaries into the stack in linear memory or any other protection mechanism for detecting buffer overflows at runtime.
This issue is aggravated by the fact that there are no guard pages between memory regions, allowing, e.g., a buffer overflow on the stack to run over static data or data on the heap.
There is also no way of marking parts of linear memory as read-only; it is always writable everywhere.
Due to these issues, WebAssembly programs can be exploited in practice, causing cross-site scripting, remote code execution, and other malicious behaviors~\cite{DBLP:conf/uss/0002KP20}, which motivates the need for tools to both discover memory-related vulnerabilities in WebAssembly binaries, and mitigate exploits at runtime.

\section{Overview and Motivating Example}
\label{sec:overview}

Our approach consists of two main components, as shown in \Cref{fig:overview}.
First, we present a novel \emph{binary-only canary instrumentation} (\Cref{sec:canaries}) that hardens WebAssembly applications by adding stack and heap canaries.
Second, we present a \emph{binary-only fuzzer for WebAssembly} (\Cref{sec:afl}).
It integrates several components into an effective and efficient end-to-end fuzzer: novel instrumentation to gather coverage information directly from WebAssembly binaries, a WebAssembly VM, and the input generation abilities of the proven AFL tool.
The remainder of this section illustrates our approach with a motivating example, subsequent sections fill in the details.

\begin{figure}[t]
  \hspace*{-2mm}
  \begin{subfigure}[t]{0.33\linewidth}
    \centering
    \includegraphics[page=1,height=1.95cm,trim=12mm 0 8mm 0]{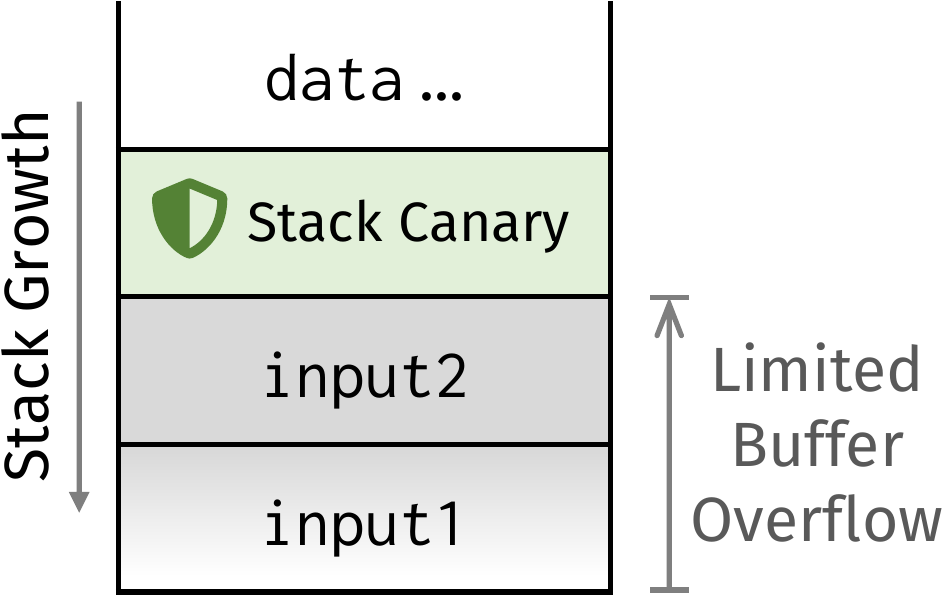}
    \captionsetup{width=.9\linewidth,justification=raggedright}
    \caption{Native, e.g., x86. Overflow is not detected but also not exploitable.}
    \label{fig:stack-native}
  \end{subfigure}
  \hfill
  \begin{subfigure}[t]{0.33\linewidth}
    \centering
    \includegraphics[page=2,height=1.95cm,trim=1mm 0 8mm 0]{figs/stack-layouts-crop.pdf}
    \captionsetup{width=.9\linewidth,justification=raggedright}
    \caption{WebAssembly, original. Overflow is exploitable and is not detected.}
    \label{fig:stack-wasm}
  \end{subfigure}
  \hfill
  \begin{subfigure}[t]{0.33\linewidth}
    \centering
    \includegraphics[page=3,height=1.95cm,trim=1mm 0 8mm 0]{figs/stack-layouts-crop.pdf}
    \captionsetup{width=.9\linewidth,justification=raggedright}
    \caption{WebAssembly, instrumented. Overflow detected, not exploitable.}
    \label{fig:stack-wasm-canary}
  \end{subfigure}
  \hspace*{-5mm}
  \vspace{-1.5mm}
  \caption{Stack layouts of the example program.}
  \label{fig:stack}
  \vspace{-2mm}
\end{figure}

\parag{Example}
The program in \Cref{fig:running-example} suffers from a textbook buffer overflow on the stack (line~3) that can be potentially triggered by the right inputs (lines~10 and 3).
Because of differences in compilers, system libraries, and protection features, the vulnerability is not exploitable when compiled to a modern native architecture, such as x86-64, but it can be exploited when compiled to WebAssembly~\cite{DBLP:conf/uss/0002KP20}.
\Cref{fig:stack-native} and \subref{fig:stack-wasm} show the stack layout of the program when executed as a native binary (compiled with GCC) and as a WebAssembly binary (compiled with Emscripten).
Because the variables \code{input1} and \code{input2} are stored in a different order on the stack, an attacker overflowing \code{input1} cannot change the program behavior in the native binary, but \emph{can} do so in WebAssembly.
It is thus important to fuzz test the WebAssembly binary, and not only a native binary compiled from the same source.

\begin{figure}[t]
  \vspace{-2mm}
\begin{lstlisting}[language=C]
void vulnerable() {
    char input1[8];
    scanf("%16s", input1);  // Buffer overflow!
    char input2[8];
    scanf("%8s", input2);
    /* more code... */  }
int read_int() {  int i; scanf("%d", &i); return i;  }
int main(int argc, char** argv) {
    char data[10] = "some data";
    if (read_int() == 42)  // Input, figured out by fuzzer.
        vulnerable();
    if (strcmp(data, "some data") == 0)
         puts("equal");
    else puts("not equal");  }
\end{lstlisting}
  \vspace{-3mm}
  \caption{Example program with a vulnerability (simplified).}
  \label{fig:running-example}
  \vspace{-2mm}
\end{figure}

\parag{Canary instrumentation}
To detect executions that exploit vulnerabilities like the above example, we present a binary-only instrumentation technique that adds protections in the form of stack and heap canaries.
The approach instruments every function in the binary with code that inserts a canary onto the stack frame upon entry and checks it upon function exit.
Beyond stack overflows, the instrumentation also detects memory violations on the heap by surrounding heap chunks with canaries.
\Cref{fig:stack-wasm-canary} shows the inserted canary on the stack of the example program.
An attack writing beyond the buffer will overwrite the canary, which the instrumented binary will detect and abort execution.

The canary instrumentation serves two purposes, marked with A) and B) in \Cref{fig:overview}.
A) The primary purpose explored in this paper is as an oracle during fuzz-testing.
If a fuzzer successfully generates an input that causes an overflow (e.g., of \code{input1} in the example), it might remain unnoticed, unless the overflow causes a crash.
Analogous to dynamic checks for memory corruptions in native programs~\cite{DBLP:conf/usenix/PrasadC03, DBLP:conf/lisa/RobertsonKMV03, Dinesh2020}, our stack and heap canary instrumentation provides a precise test oracle that warns about memory corruptions observed during an execution.
B) Beyond fuzzing, our instrumentation also serves as a hardening technique for binaries running in production.
The instrumentation mitigates exploits by detecting overflows during an execution, where it can terminate the program and hence prevent exploitation.
As we show in our evaluation, this protection comes with low overhead and be can be applied to large, real-world binaries compiled from C, C++, and Rust.

\parag{Fuzzing WebAssembly}
The second main component of our approach is the actual fuzzer.
We use a greybox fuzzing approach based on the widely used \afl{} fuzzer and its proven input generation.
The high-level workflow is that of standard greybox fuzzing:
Starting from some seed input(s), repeatedly execute the program, and gather coverage feedback,
which is used to mutate inputs until triggering a crash.
In the example program, AFL's input generation eventually figures out to start the input with \code{"42"} (line~9) to explore more behavior in function \code{vulnerable}.
However, in said function, native AFL fails to detect a vulnerability due to the different stack layouts between the native and WebAssembly binaries, whereas \name{} finds a crashing input after few minutes of fuzzing.

Applying \afl{}-style greybox fuzzing to WebAssembly is non-trivial for two reasons.
First, the fuzzer requires coverage information, which native \afl{} obtains by inserting code when compiling the program from source.
However, we want to fuzz WebAssembly binaries without requiring access to the source code.
We hence present a novel binary-only instrumentation technique to gather \afl{}-compatible coverage information from WebAssembly binaries.
Second, to be practical, fuzzers typically perform hundreds of executions of a program per second.
We present a set of WebAssembly-specific adaptions of the original \afl{} that achieve this level of efficiency.

\section{Hardening WebAssembly Programs with a Binary-Only Canary Instrumentation}
\label{sec:canaries}

The lack of virtual memory, page protections, or compiler-inserted mitigations~\cite{Cowan1998} makes WebAssembly programs more vulnerable to buffer overflows than native programs, with unique and surprising consequences such as overwriting supposedly constant data and stack-to-heap overflows~\cite{DBLP:conf/uss/0002KP20}.
While linear memory is a core part of the language that cannot be changed, over- and underflows on the stack and heap should be detected at runtime.
To do so, we present an instrumentation that statically inserts stack and heap canaries in WebAssembly binaries.
Similar to prior work on canaries on the stack~\cite{DBLP:conf/usenix/PrasadC03, DBLP:journals/access/HaJO18} or the heap~\cite{DBLP:conf/dimva/NikiforakisPJ13, DBLP:conf/lisa/RobertsonKMV03} for native programs, the basic idea is to surround memory chunks with a special value, called the \emph{canary}, and to check whether this value was overwritten, e.g., before deallocation or when returning from a function.
Our approach differs from prior work in three ways.
First, to the best of our knowledge, we are the first to present a canary-based protection for WebAssembly.
Current compilers do not implement canaries, so many already existing WebAssembly binaries are potentially vulnerable.
Second, in contrast to, e.g., compiler-inserted canaries or the popular AddressSanitizer~\cite{DBLP:conf/usenix/SerebryanyBPV12}, our approach does not require source code but instruments WebAssembly binaries directly.
This allows us to retroactively harden existing binaries.
Finally, in contrast to binary-only techniques for native programs, such as Valgrind~\cite{DBLP:conf/pldi/NethercoteS07} or Intel Pin~\cite{DBLP:conf/pldi/LukCMPKLWRH05}, our approach performs reliable, static instrumentation.
As a result, the runtime overhead of our instrumented binaries is negligible instead of the much higher overheads caused by dynamic instrumentation.

\subsection{Stack Canaries}
\label{sec:stack_canaries}

\begin{figure}
  \vspace{-5mm}
  \hfill
  \begin{subfigure}[b]{0.3\linewidth}
    \centering
    \includegraphics[page=1,width=\linewidth]{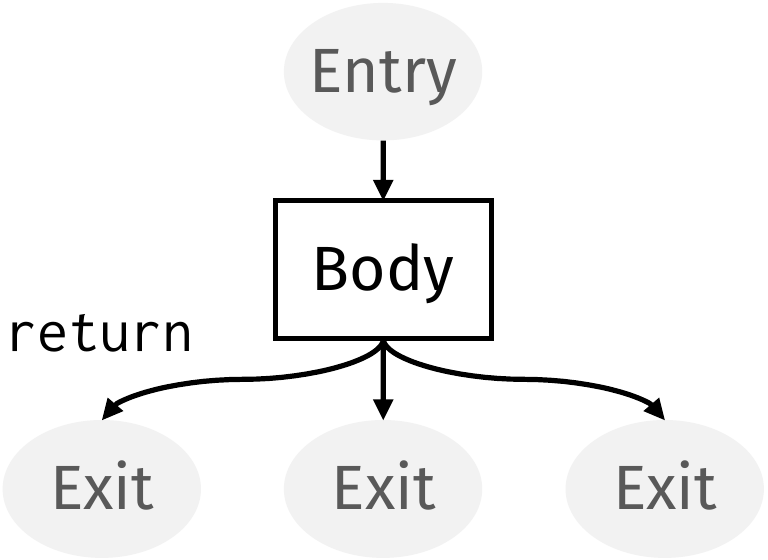}
    \caption{\raggedright{}Original body, \code{return}ing from multiple places.}
    \label{fig:cfg_without_canary_validation}
  \end{subfigure}
  \hfill
  \begin{subfigure}[b]{0.5\linewidth}
    \centering
    \includegraphics[page=2,width=.6\linewidth]{stack-canaries-crop}
    \caption{Instrumented with canaries.}
    \label{fig:cfg_with_canary_validation}
  \end{subfigure}
  \hfill
  \hspace*{0pt}
  \vspace{-1mm}
  \caption{CFGs of a WebAssembly function before and after the instrumentation. The rectangles represent subgraphs.}
  \label{fig:cfg_canary_validation}
\end{figure}

To detect buffer overflows that write beyond the current stack frame, \name{} performs three transformations on each function in a binary, as illustrated in Figure~\ref{fig:cfg_canary_validation}.
First, we insert a preamble that injects the canary value onto the stack in linear memory.
Second, we wrap the original body in a new block and rewrite all returns such that they jump to the end of said block, giving the code a single unique exit point.
Finally, we append a canary validation postamble to the function.

\newcommand{\concat}{\text{\small++}}
\begin{algorithm}
  \caption{Stack canary instrumentation.\\The \concat{} operator denotes concatenation.}
  \label{algo:stack_canary}
  \small
  \begin{algorithmic}[1]
    \Procedure {InstrumentFunction}{body}
    \State canary $\gets$ \textit{eight randomly generated bytes} \label{line:canary_generate}
    \State body $\gets$ \textsc{InjectCanary}(canary) ~\concat~ body \label{line:prepend_preamble}
    \State body $\gets \mathtt{block} ~\concat~ \text{body} ~\concat~ \mathtt{end}$ \label{line:wrap_in_block} \Comment{Wrap original body}
    \State depth $\gets$ 0 \label{line:block_depth_def}
    \For {instr in body} \label{line:body_loop_begin}
      \If {\textsc{OpensBlock}(instr)} \Comment{Track block depth}
        \State depth $\gets \text{depth} + 1$
      \ElsIf {\textsc{ClosesBlock}(instr)}
        \State depth $\gets \text{depth} - 1$
      \EndIf
      \If {instr = $\mathtt{return}$} \Comment{Redirect returns}
        \State $\text{instr} \gets \mathtt{br}\ (\text{depth} - 1)$
      \EndIf
    \EndFor \label{line:body_loop_end}
    \State $\text{body} \gets \text{body} ~\concat~ \text{\textsc{ValidateCanary}(canary)}$ \label{line:append_postamble}
    \EndProcedure
  \end{algorithmic}
\end{algorithm}

Algorithm~\ref{algo:stack_canary} presents the instrumentation of a given function in more detail.
Line~\ref{line:canary_generate} generates a random 8-byte canary value.
We use eight bytes because it is the largest primitive value supported by WebAssembly, and we use a random value to reduce the probability of missing a buffer overflow that coincidentally matches the canary.
Line~\ref{line:prepend_preamble} prepends the canary injection code to the function body.
A template of the injected code is shown in Figure~\ref{fig:canary_injection}.
The code is parametrized by the canary value \code{<CANARY>} and by \code{<SP>}, the index of the global WebAssembly variable that holds the stack pointer.
In all WASI applications, the stack pointer is the first global variable, i.e., \code{<SP>} is 0.
For non-WASI applications, heuristics to identify the stack pointer can be used~\cite{wasm-empirical}.
The code reserves space for the canary on the stack in linear memory (line~\ref{line:pre_sp_get}--\ref{line:pre_sp_set}, 16 bytes due to stack alignment) and stores the canary value there, i.e., at the beginning of the stack frame (line~\ref{line:pre_sp_get2}--\ref{line:pre_canary_store}).

The fact that a WebAssembly function may return from multiple locations raises the question where to insert code to validate the canary.
Unlike in native code, there is no single function epilogue that clears the stack and returns to the caller.
One possible approach is to separately instrument every \code{return} instruction with a copy of the validation code, but this would increase code size considerably.
Instead, we first rewrite the function to return from a single location and then insert the validation code once there.
First, the entire function is wrapped into a new WebAssembly \code{block} (line~\ref{line:wrap_in_block} of Algorithm~\ref{algo:stack_canary}). Then, each \code{return} instruction in the function body is replaced with a jump to the end of the new block (lines~\ref{line:body_loop_begin}--\ref{line:body_loop_end}), keeping the semantics of the original code.
For that, the depth variable (line~\ref{line:block_depth_def}) needs to keep track of the number of nested blocks around the current instruction.

\begin{figure}
  \vspace{-5mm}
  \begin{lstlisting}
global.get <SP>  ;; Reserve stack space for canary value.|\label{line:pre_sp_get}|
i32.const 16     ;; WASI has 16-byte stack alignment.
i32.sub
global.set <SP> |\label{line:pre_sp_set}|
global.get <SP>  ;; Store canary at beginning of stack frame.|\label{line:pre_sp_get2}|
i64.const <CANARY>
i64.store |\label{line:pre_canary_store}|
  \end{lstlisting}
  \vspace{-3mm}
  \caption{Template of \textsc{InjectCanary} for Algorithm~\ref{algo:stack_canary}.}
  \label{fig:canary_injection}
  \vspace{-2mm}
\end{figure}
\begin{figure}
  \begin{lstlisting}
block     ;; Original return value is at top of operand stack.|\label{line:post_block_begin}|
  global.get <SP>   ;; Compare canary value against reference.|\label{line:post_canary_check_begin}|
  i64.load |\label{line:post_canary_load}|
  i64.const <CANARY> |\label{line:post_canary_push}|
  i64.eq |\label{line:post_canary_compare}|
  br_if 0           ;; Jump out of block if correct. |\label{line:post_canary_check_end}|
  unreachable       ;; Otherwise: Overflow detected! |\label{line:post_canary_invalid}|
end |\label{line:post_block_end}|
global.get <SP> |\label{line:post_sp_decrement}|    ;; Adjust stack pointer.
i32.const 16
i32.add
global.set <SP> |\label{line:post_sp_set}|
return |\label{line:return}|
  \end{lstlisting}
  \vspace{-3mm}
  \caption{Template of \textsc{ValidateCanary} for Algorithm~\ref{algo:stack_canary}.}
  \label{fig:canary_validation}
  \vspace{-2mm}
\end{figure}

Finally, line~\ref{line:append_postamble} appends a postamble for validating the canary before the function returns.
Figure~\ref{fig:canary_validation} shows the template of this code.
Since all return instructions were rewritten to jump to the end of the wrapped body, the function return value will reside at the top of the WebAssembly operand stack once the postamble starts executing.
This return value is untouched by the canary validation code in lines~\ref{line:post_block_begin}--\ref{line:post_canary_check_end} of \Cref{fig:canary_validation}, avoiding the introduction of a fresh local.
The stack pointer global at this point refers to the original functions stack base, which after our injection is the memory location used for the canary.
This value is loaded from memory and compared against the known, correct canary value.
If they differ, an \code{unreachable} trap is triggered, which halts execution and thwarts potential exploits.
The trap location also indicates that a buffer overflow occurred that crosses the stack frame boundary of the function and its caller.
If the canary was intact, lines~\ref{line:post_sp_decrement}--\ref{line:post_sp_set} deallocate the canary by adjusting the stack pointer.
Finally, line~\ref{line:return} returns the only value left on the operand stack from line~\ref{line:post_block_begin}\kern-1pt: the untouched function return value.

\subsection{Heap Canaries}
\label{sec:heap_canaries}

\begin{figure}
\begin{subfigure}{.33\linewidth}
    \centering
    \includegraphics[page=2,height=1.75cm,trim=12mm 2mm 4mm 1mm]{heap-canaries-crop.pdf}
    \caption{Without canaries.}
    \label{fig:heap_without_canaries}
  \end{subfigure}
  \hfill
  \begin{subfigure}{.61\linewidth}
    \centering
    \includegraphics[page=1,height=1.75cm,trim=7mm 0 6mm 0]{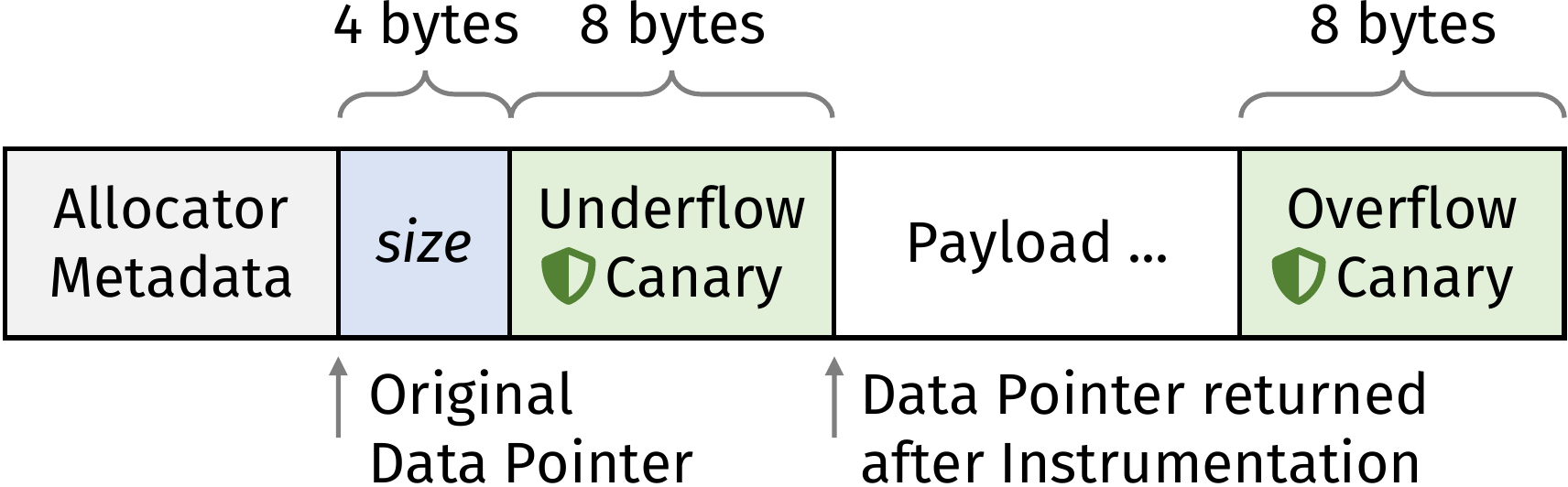}
    \caption{With canaries.}
    \label{fig:heap_with_canaries}
  \end{subfigure}
  \vspace{-1mm}
  \caption{Heap chunks, before and after the instrumentation.}
  \label{fig:heap_chunks}
  \vspace{-2mm}
\end{figure}

Detecting and preventing memory violations on the heap is important as well.
In WebAssembly, this is especially critical, because binaries frequently ship with allocators that are optimized for code size, and thus lack security features such as safe unlinking~\cite{DBLP:conf/uss/0002KP20}.
To illustrate the problem, Figure~\ref{fig:heap_without_canaries} shows the typical layout of a \emph{heap chunk}, i.e., a region of dynamically allocated memory returned by functions like \code{malloc}.
The \emph{payload} is where the user will read and write data to.
The \emph{metadata} precedes (or follows) the payload and is used for bookkeeping by the allocator.
If an attacker over- or underflows a buffer in the payload and writes into adjacent metadata, this can yield a dangerous arbitrary write primitive, which is much more powerful than a linear overflow~\cite{unsafe-unlinking1, unsafe-unlinking2}.

To detect such violations on the heap, \name{} instruments heap allocation and deallocation functions in the binary.
Our instrumentation inserts canary values before and after the payload, as illustrated by Figure~\ref{fig:heap_with_canaries}, enabling us to detect both overflows and underflows of the payload.
The canaries are inserted into the heap chunk by instrumented versions of allocation functions (\Cref{sec:heap-canary-alloc}) and checked by instrumented versions of deallocation functions (\Cref{sec:heap-canary-dealloc}).

\subsubsection{Insert Canaries on Heap Allocation}
\label{sec:heap-canary-alloc}

Heap canaries are inserted by instrumenting all functions that directly allocate heap chunks.
Our current implementation instruments the allocation functions from the C standard library, i.e., \code{malloc}, \code{calloc}, and \code{realloc}.
Other functions that allocate by calling the low-level libc functions in turn thus profit from our protection as well, as is commonly the case, e.g., for operator \code{new} in the C++ standard library.

The instrumentation inserts code into allocation functions in two places: a \emph{preamble} in the beginning and a \emph{postamble} at the end, as outlined in \Cref{algo:alloc_instrument}.
The added code has three high-level goals.
First, the allocation size needs to be increased in order to fit the canaries (line~\ref{line:increase-alloc-size}).
Second, the canary values must be written to memory (line~\ref{line:write-canaries}).
Third, the data pointer returned by the allocator needs to be adjusted before passing it to the user, such that it points to the now shifted payload (line~\ref{line:adjust-data-ptr}).
Additionally, we add two new locals to the function (lines~\ref{line:local-req-size} and \ref{line:local-data-ptr}) to save data (lines~\ref{line:save-alloc-req-size} and \ref{line:save-data-ptr}) used later by the inserted code.

Effectively, our canary code is interposed between the original allocator and the client code requesting the allocation, and must be transparent to both.
From the allocator's point of view, the payload is the whole region after the metadata in \Cref{fig:heap_with_canaries}, including the inserted canaries.
For the client code requesting the allocation, the payload is only the region between the canaries, matching the originally requested size.
Both are unaware of the data inserted by our added code.

\begin{algorithm}[t]
  \caption{Instrumentation of heap allocation functions.}
  \label{algo:alloc_instrument}
  \small
  \begin{algorithmic}[1]
    \small
    \Procedure {InstrumentAllocFunction}{f}
    \State $\text{local}_\text{req\_size} \gets \text{\textsc{AddFreshLocal}(f)}$ \label{line:local-req-size}
    \State $\text{f.body} \gets$ \Comment{Insert preamble}
    \State $\quad \text{\textsc{SaveAllocRequestSize}(f, }\text{local}_\text{req\_size}\text{)} ~\concat{}$\label{line:save-alloc-req-size}
    \State $\quad \text{\textsc{IncreaseAllocSize}(f, }\text{local}_\text{req\_size}\text{)} ~\concat{}~ \text{f.body}$\label{line:increase-alloc-size}
    \State $\text{local}_\text{data\_ptr} \gets \text{\textsc{AddFreshLocal}(f)}$\label{line:local-data-ptr}
    \State $\text{f.body} \gets \text{f.body} ~\concat{}$ \Comment{Insert postamble}
    \State $\quad \text{\textsc{SaveDataPointer}(}\text{local}_\text{data\_ptr}\text{)} ~\concat{}$\label{line:save-data-ptr}
    \State $\quad \text{\textsc{WriteSizeAndCanaries}(}\text{local}_\text{req\_size}\text{, }\text{local}_\text{data\_ptr}\text{)} ~\concat{}$\label{line:write-canaries}
    \State $\quad \text{\textsc{AdjustDataPointer}(}\text{local}_\text{data\_ptr}\text{)}$\label{line:adjust-data-ptr}
    \EndProcedure
  \end{algorithmic}
\end{algorithm}

\parag{Details}
In the preamble, we retrieve the originally requested allocation size, save it to a local for later (line~\ref{line:save-alloc-req-size}), and increase it by 20 bytes (line~\ref{line:increase-alloc-size}).
The additional 20 bytes make space for two 8-byte canaries and a 4-byte size field.
The latter is required for the checking code (\Cref{sec:heap-canary-dealloc}).
The exact inserted preamble instructions depend on the allocator function.
For \code{malloc(size\_t)} and \code{realloc(void*,size\_t)}, \textsc{SaveAllocRequestSize} simply retrieves the first (second) function argument with \code{local.get} and stores it into a new local.
\textsc{IncreaseAllocSize} is just an addition and sets the first (second) function argument to the new value.

{
\setlength{\abovedisplayskip}{3pt}
\setlength{\belowdisplayskip}{3pt}
Correctly instrumenting \code{calloc(size\_t} \code{nitems,} \code{size\_t} \code{item\_size)} is a bit more challenging.
As the allocation size is the product of both arguments and neither has to be multiple of 20, the inserted preamble changes the arguments to
\begin{align*}
\text{\code{nitems}}_{new} &= 1 \text{, and} \\
\text{\code{item\_size}}_{new} &= \text{\code{nitems}} \times \text{\code{item\_size}} + 20
\end{align*}
Additionally, \textsc{IncreaseAllocSize} checks that the second expression does not result in an integer overflow to ensure that the instrumentation never introduces errors into the program.
}

After the preamble, the original allocator code performs the regular memory allocation routine.
Then follows our inserted postamble, shown in \Cref{fig:heap_canary_injection_post}.
Since the postamble executes after the original body of the allocation function, the top-most element on the operand stack is the original return value, i.e., the pointer to the newly allocated memory.
This is saved to a local (line~\ref{line:post-store-data-ptr}).
Then, the chunk size and underflow canary are stored before the payload (lines~\ref{line:heap_post_get_size}--\ref{line:heap_post_store_underflow}).
The overflow canary is stored after the payload  (lines~\ref{line:heap_post_store_overflow_begin}--\ref{line:heap_post_store_overflow_end}).
Finally, the data pointer is fetched and adjusted to point past the underflow canary (lines~\ref{line:adjust-pointer-source-begin}--\ref{line:adjust-pointer-source-end}).
This value is finally returned to the calling code.
The result of this instrumentation is that memory allocation functions create chunks as shown in Figure~\ref{fig:heap_with_canaries}.

\begin{figure}
  \vspace{-2mm}
  \begin{lstlisting}[escapechar=@]
local.set <DATA_PTR> ;; Save pointer returned by allocator.@\label{line:post-store-data-ptr}@
local.get <DATA_PTR> ;; \ @\,\label{line:heap_post_get_size}@
local.get <REQ_SIZE> ;; @\,@| Write requested allocation size
i32.store            ;; /@\,@ at the beginning (data_ptr).
local.get <DATA_PTR> ;; \
i64.const <CANARY>   ;; @\,@| Write underflow canary
i64.store offset=4   ;; /@\,@ at data_ptr + 4. @\label{line:heap_post_store_underflow}@
local.get <DATA_PTR> ;; \ @\label{line:heap_post_store_overflow_begin}@
local.get <REQ_SIZE> ;; @\,@| Write overflow canary
i32.add              ;; @\,@| at data_ptr + size + 12.
i64.const <CANARY>   ;; @\,@|
i64.store offset=12  ;; / @\label{line:heap_post_store_overflow_end}@
local.get <DATA_PTR> ;; \ @\label{line:adjust-pointer-source-begin}@
i32.const 12         ;; @\,@| Adjust returned pointer to payload.
i32.add              ;; / @\label{line:adjust-pointer-source-end}@
  \end{lstlisting}
  \vspace{-3mm}
  \caption{Template of the inserted postamble in Algorithm~\ref{algo:alloc_instrument}.}
  \label{fig:heap_canary_injection_post}
  \vspace{-4mm}
\end{figure}

\subsubsection{Check Canaries on Heap Deallocation}
\label{sec:heap-canary-dealloc}

\name{} checks whether the heap canaries are valid whenever a heap chunk gets deallocated.
Similar to the allocation functions, the approach requires the list of deallocation functions, and our implementation currently supports the functions provided by the C standard library, i.e., \code{free} and \code{realloc}.

The heap canaries are validated by the code in Figure~\ref{fig:heap_canary_validation}, which our approach inserts at the beginning of every deallocation function.
The argument to the function is a pointer to the payload of the heap chunk.
To make the canaries transparent to the deallocation function, this pointer needs to be decremented in the beginning~(lines~\ref{line:heap_decrement_data_ptr_begin}--\ref{line:heap_decrement_data_ptr_end}).
Lines~\ref{line:heap_underflow_check_begin}--\ref{line:heap_underflow_check_end} validate the underflow canary, and lines~\ref{line:heap_overflow_check_begin}--\ref{line:heap_overflow_check_end} validate the overflow canary.
The approach uses the size stored during the canary injection to compute the location of the overflow canary.

When to check the heap canaries is a trade-off between performance, complexity of the instrumentation, and the likelihood of catching buffer overflows.
\name{} performs this check during deallocation, which is inexpensive, as every canary is checked at most once, but has the disadvantage of not catching overflows in chunks that are never deallocated.
Others have proposed more aggressive techniques that check canaries at every memory read or write~\cite{DBLP:conf/usenix/SerebryanyBPV12} or validate canaries at every syscall~\cite{DBLP:conf/dimva/NikiforakisPJ13}.
While these approaches may detect more attacks in production, they also impose a larger runtime overhead, which makes them less suited for binary hardening and fuzzing.

\begin{figure}
  \vspace{-2mm}
  \begin{lstlisting}[escapechar=@]
local.get <PARAM>    ;; The argument passed to, e.g., free().@\label{line:heap_decrement_data_ptr_begin}@
i32.const 12         ;; Adjust the pointer before passing
i32.sub              ;; it to the allocator.
local.set <PARAM> @\label{line:heap_decrement_data_ptr_end}@
block                ;; Check underflow canary.@\label{line:heap_underflow_check_begin}@
  local.get <PARAM>
  i64.load offset=4
  i64.const <CANARY>
  i64.eq
  br_if 0
  unreachable         ;; Underflow detected!
end @\label{line:heap_underflow_check_end}@
block                ;; Check overflow canary.@\label{line:heap_overflow_check_begin}@
  local.get <PARAM>  ;; \ Load payload size from our own
  i32.load           ;; / metadata at beginning of chunk.
  local.get <PARAM>  ;; \
  i32.add            ;; @\,@| Load overflow canary from
  i64.load offset=12 ;; @\,@| data_ptr + size + 12.
  i64.const <CANARY> ;; /
  i64.eq
  br_if 0
  unreachable       ;; Overflow detected!
end @\label{line:heap_overflow_check_end}@
  \end{lstlisting}
  \vspace{-3mm}
  \caption{Preamble injected into heap deallocation functions to validate heap canaries.}
  \label{fig:heap_canary_validation}
  \vspace{-4mm}
\end{figure}

\section{Binary-Only Fuzzing for WebAssembly}
\label{sec:afl}

This section presents the first binary-only fuzzer for WebAssembly.
We take a greybox fuzzing approach and build upon the popular \afl{} framework, enabling us to reuse its effective input generation abilities.
Because \afl{} usually targets programs with source code available and does not support WebAssembly, there are two key challenges to address.
The first challenge is gathering \afl{}-compatible coverage information during the execution of a WebAssembly program.
Section~\ref{sec:afl_instrumentation} describes how \name{} addresses this challenge via a novel static instrumentation of WebAssembly binaries.
The second challenge is how to integrate executions of WebAssembly on a VM with the existing \afl{} framework in a way that allows for performing hundreds of executions of a program within a second, which is the level of efficiency \afl{} provides for natively compiled code.
Section~\ref{sec:afl_port} describes how \name{} addresses this challenge through a set of novel techniques that connect \afl{} to WebAssembly.

\subsection{Coverage Instrumentation}
\label{sec:afl_instrumentation}

Greybox fuzzing is effective because it relies on lightweight feedback during program execution to steer the fuzzer.
To collect that feedback, native \afl{} compiles applications from source, inserting code to track an approximate form of path coverage~\cite{Boehme2019}, which is stored into a \emph{trace bits array}.
Unlike native \afl{}, we fuzz WebAssembly binaries without access to their source code, and hence, cannot instrument during compilation.
\afl{} also offers a QEMU mode for dynamic binary instrumentation, but it comes with a high performance overhead and naturally is architecture-specific, offering no WebAssembly implementation.
Instead, \name{} gathers coverage via static binary instrumentation that inserts code at all branches to extract \afl{}-compatible coverage information.
Hence, instrumentation cost is a one-time effort.

\begin{algorithm}
  \caption{Insertion of \afl{} coverage instrumenation.}
  \label{algo:afl}
  \small
  \begin{algorithmic}[1]
    \Procedure {afl\_instrument\_function}{$\text{f}$}
      \State $\text{depth} \gets 0$
      \State $\text{targets} \gets \{\}$
      \For {$\text{instr}$ in $\text{f.body}$}
        \If {$\text{instr} \in \{\mathtt{block}, \mathtt{if}, \mathtt{else}, \mathtt{loop}\}$}
          \If {$\text{instr} \in \{\mathtt{block}, \mathtt{if}, \mathtt{loop}\}$}
            \State $\text{depth} \gets \text{depth} + 1$ \label{line:depth_increase}
          \EndIf
          \If {$\text{instr} \in \{\mathtt{if}, \mathtt{else}, \mathtt{loop}\}$}
            \State $\text{mark(instr)}$ \label{line:mark_then}
          \EndIf
        \ElsIf {$\text{instr} = \mathtt{br\_if\ n}$} \label{line:br_if}
          \State $\text{targets} \gets \text{targets} \cup \{\text{depth} - \mathtt{n}\}$
          \State $\text{mark(instr)}$ \label{line:mark_br}
        \ElsIf {$\text{instr} = \mathtt{br\_table(}\text{jmp\_targets}\mathtt{)}$} \label{line:br_table}
          \State $\text{targets} \gets \text{targets} \cup \bigcup_{\text{t} \in \text{jmp\_targets}} \{\text{t} - \mathtt{n}\}$
        \ElsIf {$\text{instr} =  \mathtt{end}$} \label{line:end}
          \If {$\text{depth} \in \text{targets}$}
            \State $\text{mark(instr)}$ \label{line:mark_end}
            \State $\text{targets} \gets \text{targets} \setminus \{\text{depth}\}$
          \EndIf
          \State $\text{depth} \gets \text{depth} - 1$ \label{line:depth_decrease}
        \EndIf
      \EndFor
      \State$\text{mark(f.body[0])}$ \label{line:mark_function}
    \EndProcedure
  \end{algorithmic}
\end{algorithm}

As a prerequisite for instrumentation, the approach determines all branches.
The structured control flow of WebAssembly enables \name{} to precisely identify all branching points in a program.
Algorithm~\ref{algo:afl} summarizes this step, traversing each function of a binary and marking instructions that correspond to branches.
This includes $\mathtt{br\_if}$ (line~\ref{line:mark_br}), but also $\mathtt{if}$, $\mathtt{else}$ and $\mathtt{loop}$ blocks (line~\ref{line:mark_then}) since they also correspond to a branch.
Furthermore, the algorithm keeps track of the depth of each instruction, i.e., $\text{depth}$ is incremented at $\mathtt{block}$, $\mathtt{if}$ and $\mathtt{loop}$ instructions (line~\ref{line:depth_increase}) and decremented at an $\mathtt{end}$ instruction (line~\ref{line:depth_decrease}).
Keeping track of the depth is necessary to compute which $\mathtt{end}$ blocks are targets of branches.
Whenever the algorithm encounters a conditional break (either $\mathtt{br\_if}$ on line~\ref{line:br_if} or $\mathtt{br\_table}$ on line~\ref{line:br_table}), it adds the target depth of the branch instruction to the $\text{targets}$ set.
At every $\mathtt{end}$ instruction, the algorithm then checks whether the depth of that $\mathtt{end}$ instruction is in the $\text{targets}$ set (line~\ref{line:end}).
In case it is present, the $\mathtt{end}$ is a target of some branch, and is therefore also marked for instrumentation (line~\ref{line:mark_end}).
In addition to the instructions marked by Algorithm~\ref{algo:afl}, \name{} also marks the beginning of every function (line~\ref{line:mark_function}) since an indirect function call also represents a branch.

\begin{figure}
  \vspace{-2mm}
  \begin{lstlisting}[escapechar=@]
i32.const <CUR_LOCATION> @\label{line:xor_compute_begin}@         ;; Id of current branch.
global.get <PREV_LOCATION>        ;; Id of previous branch.
i32.xor @\label{line:xor_compute_end}@
global.get <TRACE_BITS> @\label{line:trace_bits_update_begin}@
i32.add
local.tee $l                      ;; Set local without pop.
local.get l
i32.load8_u                     ;; Unsigned load of 1 byte.
i32.const 1
i32.add
i32.store8                  ;; Store counter in trace_bits.@\label{line:trace_bits_update_end}@
i32.const <CUR_LOCATION @$\gg$@ 1>          ;; Shift right once.@\label{line:prev_location_update_begin}@
global.set <PREV_LOCATION> @\label{line:prev_location_update_end}@
  \end{lstlisting}
  \vspace{-2mm}
  \caption{\afl{}-style coverage instrumentation in Wasm.}
  \label{fig:afl_shim}
  \vspace{-4mm}
\end{figure}

Given the branching points identified by Algorithm~\ref{algo:alloc_instrument}, \name{} inserts instrumentation code into each of them.
A template of the instrumentation code is shown in Figure~\ref{fig:afl_shim}.
It adapts the coverage mechanism described in the AFL documentation\footnote{\url{https://lcamtuf.coredump.cx/afl/technical_details.txt}} to WebAssembly.
The basic idea is to maintain a global array of counters, the trace bits array, that indicates how often each consecutive pair of branches has been taken.
Every branch target is assigned a random identifier.
Whenever a branch is taken, the instrumentation code computes an index that combines the identifier of the current branch, \code{<CUR\_LOCATION>}, and the identifier of the previous branch, \code{<PREV\_LOCATION>} (lines~\ref{line:xor_compute_begin}--\ref{line:xor_compute_end}).
The code then increments the corresponding index of the trace bits array (lines~\ref{line:trace_bits_update_begin}--\ref{line:trace_bits_update_end}).
To initialize the trace bits array, \name{} also injects code into the \code{\_start} function of the binary, which is the WASI entry point.

\subsection{Integrating a WebAssembly VM and AFL}
\label{sec:afl_port}

The native version of \afl{} is heavily optimized towards performing as many executions of the target program within a given time period as possible.
The following presents several novel techniques that allow \name{} to achieve a similar level of efficiency.
Our implementation targets WebAssembly binaries using the WASI syscall interface, i.e., applications running on a compliant VM, such as Wasmer\footnote{\url{https://wasmer.io/}} or Wasmtime.

\parag{Avoiding VM restarts}
With the fuzzed program running on a VM, one possible approach is to start a new instance of the VM for each run of the target program.
However, doing so may easily result in more time spend on starting the VM and compiling the module to native code than on running the target program.
Instead, \name{} starts the VM once, lets the VM precompile the target WebAssembly binary, and then reuses both throughout the fuzzing process.
\name{} uses the Wasmtime C API\footnote{\url{https://docs.wasmtime.dev/c-api/}} to separately compile, then instantiate, and finally run WebAssembly modules.
For every newly generated input, the fuzzer instantiates the WebAssembly module that was already compiled to native code and then calls the \code{\_start} function, i.e., the entry point of the target binary.

\parag{Accessing the trace bits array}
To generate new inputs, \afl{} needs to access the coverage information stored in the trace bits array.
The native version of \afl{} starts the target program as a subprocess and accesses the trace bits array via shared memory.
Instead, \name{} exploits the fact that the VM sandbox allows for the target program and AFL's own code to share a single address space.
To this end, our approach inserts an accessor function into the binary during the coverage instrumentation.
This function returns a pointer to the trace bits array in the linear memory of the target program.
After each execution of the target program, \name{} calls the special function and extracts the 64KB of linear memory, corresponding to the trace bits, using the Wasmtime C API.

\parag{Detecting crashes}
The native version of \afl{} detects crashes by looking for fatal signals (SIGSEGV, SIGKILL, SIGABRT) in the target program.
However, WASI does not support signals, so \name{} uses the trap system of WebAssembly to determine when the target program crashes.
To this end, the oracles that \name{} inserts (Section~\ref{sec:canaries}) trigger an \texttt{unreachable} trap when they detect an overflow or underflow.
In addition, WebAssembly has other built-in traps that also indicate faulty behavior (Section~\ref{sec:background}).
When the \code{\_start} function terminates, \name{} checks if the termination was triggered by a trap, and in that case, marks it as a crash.\footnote{Programs terminating with a non-zero exit code also trigger a trap in WASI, but \name{} takes care to not interpret these specific traps as crashes, as they do not indicate faulty behavior as crashes in native programs do.}

\parag{Killing long-running executions}
Randomly generated inputs may trigger long-running or even non-terminating executions.
To prevent those from slowing down overall fuzzing, native \afl{} runs the fuzzed program in a separate process, which is killed after a timeout.
However, since the WebAssembly VM and generated code are running in the same process as \afl{}, \name{} implements two mechanisms to stop long-running executions.
First, it uses a ``soft killing'' mechanism based on a separate thread running on the WASI VM that interrupts the thread of the target program after a timeout dynamically set by \afl{}.
Second, to address that the target program may not react to the interrupt, a second ``hard killing'' mechanism may restart the entire VM after a longer timeout.

\section{Evaluation}
\label{sec:evaluation}

We evaluate \name{} along the two use-cases we present in \Cref{sec:overview}.
First, \emph{end-to-end fuzzing} of WebAssembly binaries:
\begin{itemize}[itemsep=-2pt, topsep=6pt]
  \item[\textbf{RQ1}] \emph{Effectiveness}:
  How effective is \name{} at covering paths and finding crashes?
\item[\textbf{RQ2}] \emph{Robustness}:
  How robust is the instrumentation when applied to real-world binaries?
  \item[\textbf{RQ3}] \emph{Efficiency}:
  How efficient is fuzzing with \name{}?
\end{itemize}

\noindent
Second, \emph{hardening binaries for production} through canaries:
\begin{itemize}[itemsep=-2pt, topsep=6pt]
  \item[\textbf{RQ4}] \emph{Effectiveness}:
  How effective are the inserted canaries at preventing previously demonstrated exploits?
  \item[\textbf{RQ5}] \emph{Efficiency}:
  How much overhead do the canaries impose?
\end{itemize}

\noindent
For reproducibility, future research, and practicioners, we make our source code, data, and all experimental results available at \url{https://github.com/fuzzm/fuzzm-project}.

\subsection{Experimental Setup}
\label{sec:benchmarks}

\parag{Benchmarks}
We use three sets of benchmarks (Table~\ref{table:fuzzing_results}).
Benchmarks~1 to~7 are real-world applications and libraries that can be compiled to WebAssembly with WASI.
The selected versions of those programs suffer from known memory vulnerabilities, which a fuzzer might help to uncover and fix.

Benchmarks~8 to~10 are from the \mbox{LAVA-M} benchmark~\cite{DBLP:conf/sp/Dolan-GavittHKL16}.
We omit the \textsl{who} program in \mbox{LAVA-M} since it reads the list of mounted file systems, which is not yet supported by WASI.
\afl{}, and by extension also \name{}, is known to perform poorly on LAVA-M as the fuzzer does not handle the multi-byte constraints of \mbox{LAVA-M} bugs well~\cite{rawat2017vuzzer}.
Because \mbox{LAVA-M} has been criticized for not being representative of real bugs~\cite{DBLP:conf/ccs/KleesRCW018}, we would have liked to instead evaluate against the more modern Magma benchmark suite~\cite{DBLP:journals/corr/abs-2009-01120}.
However, all the Magma benchmarks use features not yet supported by WASI, such as threads, networking, and long jumps.

Benchmarks~11 to~28 are real-world WebAssembly binaries, gathered from public websites, GitHub, and NPM packages by the WasmBench dataset~\cite{wasm-empirical}.
We select 18 binaries that run without error (before any instrumentation) in the Wasmtime VM.
Among them are large applications, such as SQLite and Clang compiled to WebAssembly, but also several smaller binaries, such as a JSON formatter (\emph{canonicaljson}), a template engine (\emph{handlebars-cli}), and an interpreter (\emph{bfi}).

\parag{Compilation}
We compile the source code from the first and second set using a version of Clang that targets WebAssembly\footnote{\url{https://github.com/WebAssembly/wasi-sdk}} and then instrument the binaries as described in Sections~\ref{sec:canaries} and~\ref{sec:afl}.
For comparing against native \afl{}, we compile the benchmarks with the \afl{} version of GCC.
Since AFL's instrumentation is applied during compilation, this is not completely true to our scenario where source code is not available and binaries are compiled for production.
To level the ground, we do not use AddressSanitizer or any other oracle that requires source code, neither for \name{} nor \afl{}.
An alternative baseline would be the QEMU mode of \afl{}, which uses dynamic instrumentation, but since it is much slower than normal \afl{}, it would give \name{} an unfair advantage.
For the third benchmark set, we do not have any source code, which highlights the need for a binary-only fuzzer.

\parag{Repetitions and system configuration}
We repeatedly fuzz each benchmark five times for 24 hours, both with \name{} and \afl{}.
The repetitions address the variance of results due to the inherent non-determinism of fuzzing~\cite{DBLP:conf/ccs/KleesRCW018}.
In addition to the mean results across the repetitions, we report 95\% confidence intervals.
All experiments were performed on two machines, each with two Intel Xeon 12-core 24-thread CPUs running at 2.2\,GHz, using 256\,GB of system memory, and Ubuntu 18.04 LTS.
For AFL, we use version~2.57b.

\subsection{RQ1: Effectiveness of \name{}}
\label{sec:rq-fuzzing-effectiveness}

\begin{table*}[t]
  \vspace{-4mm}
  \caption{Benchmarks overview and fuzzing results (5$\,\times\,$24 hours). The reported numbers are mean and 95\% confidence intervals.}
  \label{table:fuzzing_results}
  \vspace*{-1mm}
  \setlength\tabcolsep{4pt}
\footnotesize
\begin{adjustbox}{center}
\begin{tabular}{r l @{\qquad} rCr rCr rCr rCr rCr rCr rCr rCr}
  \toprule
  && \multicolumn{15}{c}{\name{} (WebAssembly binaries)} & \multicolumn{9}{c}{\afl{} (native, built from source)} \\
  \cmidrule(lr){3-17}
  \cmidrule(lr){18-26}
  \# & Benchmark & \multicolumn{3}{c}{Paths} & \multicolumn{9}{c}{Crashes, of those: caused by Canaries} & \multicolumn{3}{c}{Execs/sec}  & \multicolumn{3}{c}{Paths} & \multicolumn{3}{c}{Crashes} & \multicolumn{3}{c}{Execs/sec} \\
  \cmidrule(lr){6-14}
  &&&&& \multicolumn{3}{c}{Total} & \multicolumn{3}{c}{Stack Can.} & \multicolumn{3}{c}{Heap Can.} \\
  \midrule

  \multicolumn{26}{@{}l}{\emph{Benchmark set 1 -- Real-world applications and libraries:}}\\[2pt]
   1 & \texttt{abc2mtex}         & 1678.6 & $\pm$ & 22.5   & 267.9 & $\pm$ & 6.6 & 224.5 & $\pm$ & 5.5 & 4.2 & $\pm$ & 1.6   & 359.4 & $\pm$ & 42.4    & 2999.2 & $\pm$ & 82.6  & 820.1 & $\pm$ & 63.2 & 879.6 & $\pm$ & 212.1   \\
   2 & \texttt{flac}             & 607.5 & $\pm$ & 11.5    & 0.0 & $\pm$ & 0.0   & 0.0 & $\pm$ & 0.0   & 0.0 & $\pm$ & 0.0   & 497.0 & $\pm$ & 7.1     & 1617.1 & $\pm$ & 75.1  & 0.0 & $\pm$ & 0.0    & 1228.1 & $\pm$ & 391.3  \\
   3 & \texttt{jbig2dec}         & 2199.1 & $\pm$ & 13.8   & 0.1 & $\pm$ & 0.2   & 0.0 & $\pm$ & 0.0   & 0.0 & $\pm$ & 0.0   & 66.2 & $\pm$ & 51.6     & 3330.1 & $\pm$ & 49.8  & 0.0 & $\pm$ & 0.0    & 437.7 & $\pm$ & 264.8   \\
   4 & \texttt{libpng}           & 727.0 & $\pm$ & 10.4    & 96.1 & $\pm$ & 4.0  & 0.0 & $\pm$ & 0.0   & 77.2 & $\pm$ & 3.8  & 430.9 & $\pm$ & 67.6    & 1123.4 & $\pm$ & 25.3  & 176.4 & $\pm$ & 2.8  & 692.5 & $\pm$ & 481.6   \\
   5 & \texttt{libtiff}          & 860.3  & $\pm$ & 9.1    & 0.0   & $\pm$ & 0.0 & 0.0 & $\pm$ & 0.0   & 0.0 & $\pm$ & 0.0   & 868.5 & $\pm$ & 64.3    & 2542.5 & $\pm$ & 33.6  & 0.0    & $\pm$ & 0.0 & 953.7  & $\pm$ & 467.8  \\
   6 & \texttt{openjpeg}         & 5322.2 & $\pm$ & 3611.0 & 90.3 & $\pm$ & 39.3 & 7.7 & $\pm$ & 4.5   & 8.6 & $\pm$ & 10.4  & 457.3 & $\pm$ & 242.8   & 1779.7 & $\pm$ & 39.8  & 90.7 & $\pm$ & 3.4   & 605.4 & $\pm$ & 435.4   \\
   7 & \texttt{pdfresurrect}     & 840.1 & $\pm$ & 207.0   & 54.5 & $\pm$ & 8.4  & 15.1 & $\pm$ & 3.4  & 17.2 & $\pm$ & 5.4  & 228.1 & $\pm$ & 194.1   & 1011.0 & $\pm$ & 226.3 & 129.9 & $\pm$ & 29.2 & 701.5 & $\pm$ & 369.2   \\[2pt]

  \multicolumn{26}{@{}l}{\emph{Benchmark set 2 -- From LAVA-M~\cite{DBLP:conf/sp/Dolan-GavittHKL16}:}}\\[2pt]
   8 & \texttt{base64}          & 200.6  & $\pm$ & 7.2    & 34.4  & $\pm$ & 1.5  & 0.0   & $\pm$ & 0.0 & 0.0   & $\pm$ & 0.0 & 225.8 & $\pm$ & 83.7  & 355.8  & $\pm$ & 29.1  & 0.1    & $\pm$ & 0.2  & 514.3  & $\pm$ & 276.6 \\
   9 & \texttt{md5sum}          & 395.2  & $\pm$ & 20.6   & 0.0   & $\pm$ & 0.0  & 0.0   & $\pm$ & 0.0 & 0.0   & $\pm$ & 0.0 & 324.6 & $\pm$ & 202.7 & 317.9  & $\pm$ & 8.9   & 0.0    & $\pm$ & 0.0  & 202.4  & $\pm$ & 60.7  \\
  10 & \texttt{uniq}            & 213.1  & $\pm$ & 22.8   & 0.0   & $\pm$ & 0.0  & 0.0   & $\pm$ & 0.0 & 0.0   & $\pm$ & 0.0 & 678.2 & $\pm$ & 109.6 & 113.0  & $\pm$ & 3.7   & 0.3    & $\pm$ & 0.4  & 415.0  & $\pm$ & 337.5 \\[2pt]

  \multicolumn{26}{@{}l}{\emph{Benchmark set 3 -- Real-world WebAssembly binaries from WasmBench~\cite{wasm-empirical}:}}\\[2pt]
  11 & \texttt{bf}               & 271.4 & $\pm$ & 27.5    & 0.0 & $\pm$ & 0.0   & 0.0 & $\pm$ & 0.0   & 0.0 & $\pm$ & 0.0   & 28.6 & $\pm$ & 7.4    & \multicolumn{9}{c}{\multirow{17}{*}{\shortstack{\textcolor{lightgray}{\huge{}N/A}\\[7pt] (As those samples are binary-only \\WebAssembly programs, there is no\\native counterpart to fuzz with AFL.)}}}\\
  12 & \texttt{bfi}              & 2158.0 & $\pm$ & 108.8  & 97.8 & $\pm$ & 26.1 & 0.0 & $\pm$ & 0.0   & 30.8 & $\pm$ & 12.5 & 286.4 & $\pm$ & 40.1  & \multicolumn{9}{c}{}\\
  13 & \texttt{canonicaljson}    & 357.4 & $\pm$ & 15.8    & 180.4 & $\pm$ & 6.9 & 0.0 & $\pm$ & 0.0   & 0.0 & $\pm$ & 0.0   & 428.2 & $\pm$ & 193.4 & \multicolumn{9}{c}{}\\
  14 & \texttt{clang}            & 6.0 & $\pm$ & 0.6       & 0.0 & $\pm$ & 0.0   & 0.0 & $\pm$ & 0.0   & 0.0 & $\pm$ & 0.0   & 36.8 & $\pm$ & 0.7    & \multicolumn{9}{c}{}\\
  15 & \texttt{colcrt}           & 231.0 & $\pm$ & 7.6     & 0.0 & $\pm$ & 0.0   & 0.0 & $\pm$ & 0.0   & 0.0 & $\pm$ & 0.0   & 83.6 & $\pm$ & 50.3   & \multicolumn{9}{c}{}\\
  16 & \texttt{handlebars-cli}   & 882.2 & $\pm$ & 69.5    & 39.4 & $\pm$ & 0.7  & 0.0 & $\pm$ & 0.0   & 39.4 & $\pm$ & 0.7  & 222.0 & $\pm$ & 136.8 & \multicolumn{9}{c}{}\\
  17 & \texttt{hq9\_plus\_rs}    & 227.0 & $\pm$ & 15.5    & 42.8 & $\pm$ & 0.9  & 0.0 & $\pm$ & 0.0   & 42.8 & $\pm$ & 0.9  & 111.0 & $\pm$ & 42.0  & \multicolumn{9}{c}{}\\
  18 & \texttt{jq}               & 1.0 & $\pm$ & 0.0       & 0.0 & $\pm$ & 0.0   & 0.0 & $\pm$ & 0.0   & 0.0 & $\pm$ & 0.0   & 334.0 & $\pm$ & 7.3   & \multicolumn{9}{c}{}\\
  19 & \texttt{libxml2}          & 177.8 & $\pm$ & 5.0     & 0.0 & $\pm$ & 0.0   & 0.0 & $\pm$ & 0.0   & 0.0 & $\pm$ & 0.0   & 70.0 & $\pm$ & 1.5    & \multicolumn{9}{c}{}\\
  20 & \texttt{qjs}              & 9640.0 & $\pm$ & 96.0   & 140.4 & $\pm$ & 61.4& 3.4 & $\pm$ & 3.0   & 11.6 & $\pm$ & 8.6  & 387.0 & $\pm$ & 20.0  & \multicolumn{9}{c}{}\\
  21 & \texttt{qr2text}          & 1.0 & $\pm$ & 0.0       & 0.0 & $\pm$ & 0.0   & 0.0 & $\pm$ & 0.0   & 0.0 & $\pm$ & 0.0   & 84.6 & $\pm$ & 0.4    & \multicolumn{9}{c}{}\\
  22 & \texttt{rev}              & 161.4 & $\pm$ & 6.1     & 0.0 & $\pm$ & 0.0   & 0.0 & $\pm$ & 0.0   & 0.0 & $\pm$ & 0.0   & 140.4 & $\pm$ & 20.1  & \multicolumn{9}{c}{}\\
  23 & \texttt{save}             & 23.2 & $\pm$ & 0.7      & 0.0 & $\pm$ & 0.0   & 0.0 & $\pm$ & 0.0   & 0.0 & $\pm$ & 0.0   & 46.4 & $\pm$ & 1.2    & \multicolumn{9}{c}{}\\
  24 & \texttt{sqlite}           & 4284.2 & $\pm$ & 738.3  & 2.4 & $\pm$ & 4.2   & 0.0 & $\pm$ & 0.0   & 0.0 & $\pm$ & 0.0   & 359.4 & $\pm$ & 116.5 & \multicolumn{9}{c}{}\\
  25 & \texttt{viu}              & 12.4 & $\pm$ & 1.8      & 0.0 & $\pm$ & 0.0   & 0.0 & $\pm$ & 0.0   & 0.0 & $\pm$ & 0.0   & 707.2 & $\pm$ & 7.6   & \multicolumn{9}{c}{}\\
  26 & \texttt{wasi-example}     & 213.4 & $\pm$ & 7.1     & 50.2 & $\pm$ & 0.9  & 0.0 & $\pm$ & 0.0   & 0.0 & $\pm$ & 0.0   & 764.4 & $\pm$ & 81.8  & \multicolumn{9}{c}{}\\
  27 & \texttt{wasm-interface}   & 5.2 & $\pm$ & 0.4       & 0.0 & $\pm$ & 0.0   & 0.0 & $\pm$ & 0.0   & 0.0 & $\pm$ & 0.0   & 622.2 & $\pm$ & 10.3  & \multicolumn{9}{c}{}\\
  28 & \texttt{zxing\_barcode}   & 2799.2 & $\pm$ & 236.2  & 32.2 & $\pm$ & 5.6  & 0.0 & $\pm$ & 0.0   & 17.4 & $\pm$ & 5.2  & 131.4 & $\pm$ & 62.2  & \multicolumn{9}{c}{}\\

  \midrule
  \multicolumn{2}{l}{Average (only sets 1 and 2)} & 1304.4 &       &        & 54.3 &       &      & 24.7 &       &     & 10.7  &       &      & 413.6  &       &       & 1518.9 & & & 121.7 & & & 663.0 \\
  \multicolumn{2}{l}{Average (all)}   & 1232.0 &       &        & 40.3  &       &      & 9.0 &       &     & 8.9  &       &      & 320.7 \\

  \bottomrule
\end{tabular}
\end{adjustbox}
\vspace{-3mm}
\end{table*}

We evaluate the end-to-end effectiveness of fuzzing WebAssembly binaries with \name{} by measuring how many unique paths are explored, how many unique crashes are triggered, and whether the canary instrumentation helps in finding those crashes.
Table~\ref{table:fuzzing_results} gives the results, where the numbers for \name{} are presented in the left block.
Crashes and paths are counted using \afl{}'s notion of unique crashes and unique paths, respectively, i.e., two crashes are merged if they are found on the same path.
Different crashes as per this metric may sometimes have the same root cause~\cite{DBLP:conf/ccs/KleesRCW018}.

We find that \name{} successfully explores many hundreds of paths through the programs, on average 1,232 unique paths per benchmark after 24 hours of fuzzing.
We see that this works even for complex programs such as \textsl{flac} (benchmark set~1) or \textsl{sqlite} (set~3).
For benchmark sets one and two, where the fuzzed programs are known, we provide the fuzzer with seed inputs.
For set three, we only provided an empty file as seed input, which could explain the lower average number of discovered paths.
In terms of crashes, \name{} finds 40.3 crashes per benchmark on average.
For example, for \emph{libpng} and \emph{pdfresurrect}, \name{} generates crashing inputs that produce the exact stack trace of proof-of-concept exploits against the vulnerabilities, confirming that \name{} can find real bugs.

\begin{figure}
\begin{subfigure}[b]{0.45\linewidth}
    \centering
    \includegraphics[width=\linewidth]{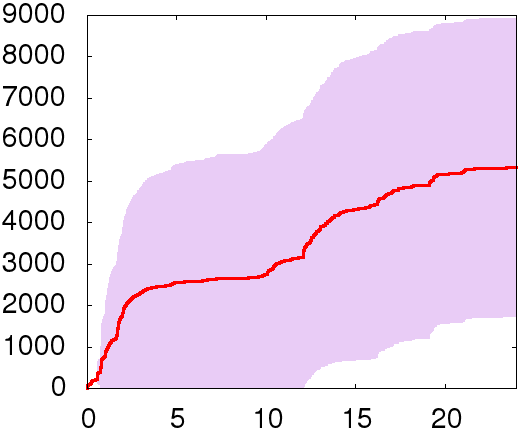}
    \vspace*{-5mm}
    \caption{\texttt{openjpeg}.}
    \label{fig:paths-over-time-openjpeg}
  \end{subfigure}
  \hfill
  \begin{subfigure}[b]{0.45\linewidth}
    \centering
    \includegraphics[width=\linewidth]{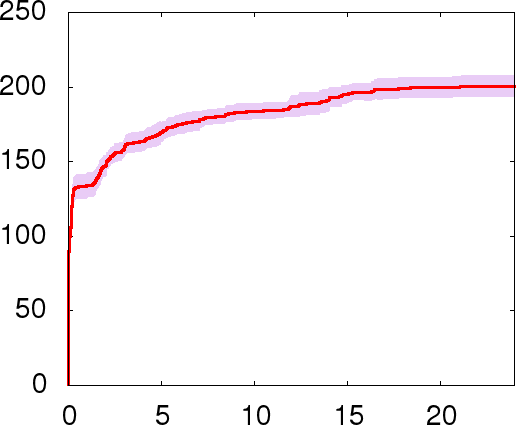}
    \vspace*{-5mm}
    \caption{\texttt{base64}.}
    \label{fig:paths-over-time-base64}
  \end{subfigure}
  \\[5pt]
  \begin{subfigure}[b]{0.45\linewidth}
    \centering
    \includegraphics[width=\linewidth]{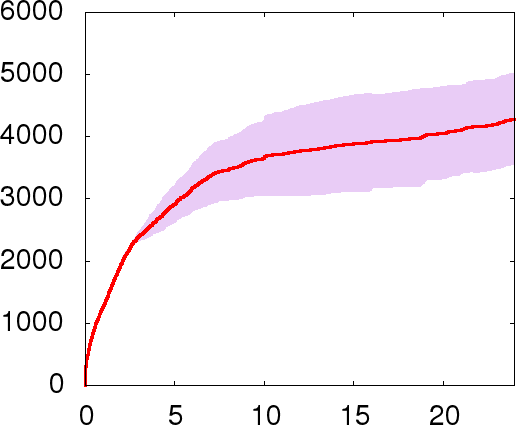}
    \vspace*{-5mm}
    \caption{\texttt{sqlite}.}
    \label{fig:paths-over-time-sqlite}
  \end{subfigure}
  \hfill
  \begin{subfigure}[b]{0.45\linewidth}
    \centering
    \includegraphics[width=\linewidth]{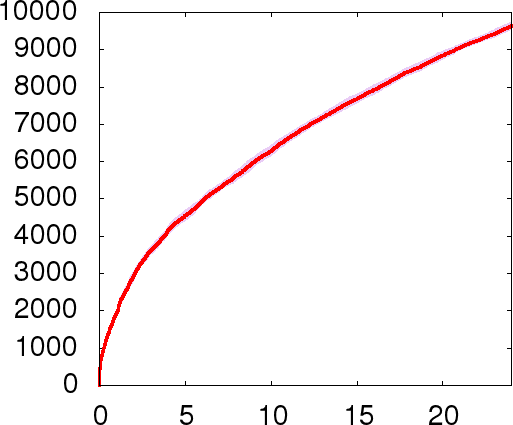}
    \vspace*{-5mm}
    \caption{\texttt{qjs}.}
    \label{fig:paths-over-time-qjs}
  \end{subfigure}
  \vspace*{-1mm}
  \caption{Average unique discovered paths (y-axis) over 24~hours of fuzzing (x-axis), with 95\% confidence intervals.}
  \label{fig:paths-over-time}
  \vspace{-3mm}
\end{figure}

To better understand how \name{} exercises a program over a 24-hour period of fuzzing, Figure~\ref{fig:paths-over-time} shows the number of unique paths detected over time, for four programs.
As the plots for found crashes strongly correlate with explored paths, we omit the former for space reasons.
(Both plots for all programs are available online.)
As is typical for fuzzers, the majority of behaviors are detected early on, usually within the first couple of hours (\ref{fig:paths-over-time-openjpeg}/\subref{fig:paths-over-time-base64}/\subref{fig:paths-over-time-sqlite}).
Then, the number of new paths and crashes often saturates, especially for the LAVA-M benchmarks (\subref{fig:paths-over-time-base64}).
For some benchmarks, e.g., \textsl{qjs}, \name{} still finds new paths when given more time (\subref{fig:paths-over-time-qjs}).
The confidence intervals are generally small, except for \textsl{openjpeg} (\subref{fig:paths-over-time-openjpeg}) and \textsl{pdfresurrect}, where the results vary considerably across runs.
Overall, these findings are consistent with previous work, and show that running a fuzzer multiple times is important to obtain statistically meaningful results~\cite{DBLP:conf/ccs/KleesRCW018}.

\parag{Comparison}
As \name{} is the first binary-only fuzzing approach for WebAssembly, we cannot directly compare to any baseline.
However, to put the number of paths and crashes into perspective, we also present results for native AFL on the right side of \Cref{table:fuzzing_results}.
This is only meant as a rough frame of reference, as a fair comparison is impossible for several reasons.
First, \name{} requires only the binary as input, whereas AFL applies its instrumentation during compilation from source.
Second, our notion of branches may differ from branches considered by AFL, simply due to different compilers and their target-dependent optimizations and codegen.
Third, unlike in native binaries, all libraries (including \emph{libc}) are statically linked in WebAssembly, which increases the amount of code \name{} instruments and thus has to fuzz.
Fourth, the number of explored paths naturally depends on the execution speed, which is in principle lower on WebAssembly compared to native (see also \Cref{sec:rq-efficiency-fuzzing}).
Finally, benchmark set three is only available as WebAssembly binaries, which is why we cannot compare against AFL for these benchmarks.

From the data on benchmark sets one and two, we can see that \name{} discovers on average a similar number of paths compared with AFL (1304 vs. 1519).
In terms of crashes, AFL triggers 122 on average, which is roughly twice as many as \name{}'s 54.
One outlier is \textsl{base64} where \name{} triggers 34 unique crashes but \afl{} triggers only one crash in one of the performed runs.
The 34 WebAssembly crashes are triggered by a built-in sanity check of the executing VM, which is not present in native binaries and explains why these crashes are not detected by \afl{}.
For programs where \name{} does not find any crashes (e.g., \textsl{flac}), AFL does not either.

For the LAVA-M benchmarks, both \name{} and \afl{} fail to trigger any of the bugs injected by the LAVA tool.
This observation is surprising since other papers that compare with \afl{} report at least some bugs detected for \textsl{uniq} and also sometimes for \textsl{base64}~\cite{DBLP:conf/cis/ZhangYFT17, DBLP:conf/sp/ChenC18}.
Manually investigating some of the LAVA-M bugs shows that they resemble use-after-free bugs and that a crash depends on the memory allocator allocating a new chunk at the exact location of some previously freed chunk.
We attribute the fact that neither \name{} nor \afl{} finds these bugs to differences in (versions of) the used memory allocator and to differences across versions of \afl{}.

\parag{Effectiveness of Canaries}
Besides being the first approach for fuzzing WebAssembly binaries, \name{} contributes canary-based oracles to detect stack and heap over- and underflows.
We measure how much these oracles contribute to the crashes detected by \name{} by distinguishing crashes caused by the oracles from other crashes.
The three ``Crashes'' columns of Table~\ref{table:fuzzing_results} show the results.
The stack and heap canaries are responsible for 22.2\% and 22.0\% of all detected crashes, respectively, on all benchmarks, and 45.5\% and 19.7\% on benchmark sets one and two.
This indicates that both contribute significantly to the effectiveness of \name{}.

\begin{result}
  Aplied to well-known applications, libraries, and real-world WebAssembly binaries, \name{} triggers an average of 40 unique crashes and 1,232 unique paths within 24 hours of fuzzing, which are similar results as AFL applied to native programs.
Our canary-based oracles detect about half of all detected crashes, and hence, contribute significantly to the effectiveness of \name{}.
\end{result}
\vspace{-2mm}

\begin{table*}[t]
  \caption{Robustness and runtime overhead of instrumented binaries (mean over 25 repetitions, 95\% confidence intervals).}  \label{table:instrumentation_results}
  \vspace*{-1mm}
  \footnotesize
  \begin{adjustbox}{center}
  \begin{tabular}{r l r @{\hspace*{25pt}} r l rCl rCl rCl rCl rCl}
    \toprule
    \multirow{2}[2]{*}{\#} & \multirow{2}[2]{*}{Benchmark} & \multirow{2}[2]{*}{\shortstack{Test\\Inputs}} & \multirow{2}[2]{*}{\shortstack{Execution Time (ms),\\Uninstrumented}\hspace*{-15pt}} & \hspace*{15pt} & \multicolumn{15}{c}{Execution Time, relative to Uninstrumented Binary} \\
    \cmidrule(lr){6-20}
    & & & & & \multicolumn{3}{c}{Coverage}& \multicolumn{3}{c}{Stack Canaries} & \multicolumn{3}{c}{Heap Canaries} & \multicolumn{3}{c}{All Canaries} & \multicolumn{3}{c}{Cov. + Can.} \\
    \midrule
    1  & \texttt{abc2mtex}      & 30     & 815    & & 1.38 & $\pm$ & 0.03 & 1.02 & $\pm$ & 0.01 & 1.02 & $\pm$ & 0.01 & 1.06 & $\pm$ & 0.01 & 1.38 & $\pm$ & 0.04 \\
    2  & \texttt{flac}          & 10     & 2,449  & & 1.42 & $\pm$ & 0.01 & 1.02 & $\pm$ & 0.01 & 1.00 & $\pm$ & 0.01 & 1.02 & $\pm$ & 0.01 & 1.48 & $\pm$ & 0.02 \\
    3  & \texttt{jbig2dec}      & 28     & 4,742  & & 2.05 & $\pm$ & 0.01 & 1.22 & $\pm$ & 0.01 & 1.00 & $\pm$ & 0.00 & 1.22 & $\pm$ & 0.01 & 2.24 & $\pm$ & 0.01 \\
    4  & \texttt{libpng}        & 10     & 3,480  & & 1.57 & $\pm$ & 0.02 & 1.03 & $\pm$ & 0.01 & 1.00 & $\pm$ & 0.01 & 1.02 & $\pm$ & 0.01 & 1.58 & $\pm$ & 0.02 \\
    5  & \texttt{libtiff}       & 10     & 899    & & 1.30 & $\pm$ & 0.03 & 1.13 & $\pm$ & 0.01 & 1.11 & $\pm$ & 0.01 & 1.14 & $\pm$ & 0.01 & 1.40 & $\pm$ & 0.03 \\
    6  & \texttt{openjpeg}      & 10     & 4,750  & & 1.77 & $\pm$ & 0.02 & 1.05 & $\pm$ & 0.01 & 1.00 & $\pm$ & 0.01 & 1.06 & $\pm$ & 0.01 & 1.84 & $\pm$ & 0.02 \\
    7  & \texttt{pdfresurrect}  & 10     & 2,894  & & 1.16 & $\pm$ & 0.02 & 1.06 & $\pm$ & 0.01 & 1.31 & $\pm$ & 0.01 & 1.35 & $\pm$ & 0.01 & 1.53 & $\pm$ & 0.02 \\
    8  & \texttt{base64}        & 10     & 256    & & 1.32 & $\pm$ & 0.10 & 1.02 & $\pm$ & 0.02 & 0.99 & $\pm$ & 0.01 & 1.03 & $\pm$ & 0.02 & 1.31 & $\pm$ & 0.09 \\
    9  & \texttt{md5sum}        & 10     & 272    & & 1.33 & $\pm$ & 0.09 & 1.03 & $\pm$ & 0.03 & 1.00 & $\pm$ & 0.01 & 1.05 & $\pm$ & 0.02 & 1.30 & $\pm$ & 0.09 \\
    10 & \texttt{uniq}          & 10     & 260    & & 1.34 & $\pm$ & 0.09 & 1.04 & $\pm$ & 0.03 & 1.02 & $\pm$ & 0.01 & 1.11 & $\pm$ & 0.09 & 1.39 & $\pm$ & 0.10 \\
    \midrule
    \multicolumn{2}{l}{Average} &        & 2,082  & & 1.46 & $\pm$ & 0.04 & 1.06 & $\pm$ & 0.02 & 1.05 & $\pm$ & 0.01 & 1.11 & $\pm$ & 0.02 & 1.54 & $\pm$ & 0.04 \\
    \bottomrule
  \end{tabular}
  \end{adjustbox}
  \vspace{-3mm}
\end{table*}

\subsection{RQ2: Robustness of Instrumentation}
\label{sec:rq2}

To effectively fuzz a program, our instrumentation should not affect the semantics of the program, except in the presence of overflows, where the canaries should terminate the program.
To validate the robustness of \name{}'s instrumentation, we compare the output generated by the non-instrumented and the instrumented versions of the benchmarks.
For each benchmark in the first two sets, we collect at least ten different inputs, totaling 138 test cases (Table~\ref{table:instrumentation_results}).
We sample these inputs from different websites\footnote{E.g., \url{https://filesamples.com/}.}, and for programs where we could not find sufficiently many examples online, e.g., \textsl{pal2rgb}, we generate inputs by, e.g., converting images to the pal format.
As benchmark set three is only available in binary form without source code or documentation, we do not generate test inputs for those programs.
For the binaries and test inputs shown in \Cref{table:instrumentation_results}, we verify that the outputs produced by the instrumented binaries are equivalent to the outputs produced by the uninstrumented binaries for all 138 test cases.
As additional evidence for the robustness of our instrumentation, we find that all binaries we instrumented pass built-in WebAssembly validation, which performs, e.g., type-checking of instructions and functions.

\begin{result}
Test runs of the benchmarks and the static validation applied to each WebAssembly module before its execution show that the binary instrumentation applied by \name{} preserves the semantics of the original program.
\end{result}
\vspace{-1mm}

\subsection{RQ3: Efficiency of End-to-End Fuzzing}
\label{sec:rq-efficiency-fuzzing}

Effective fuzzing requires many repeated executions of the target program in limited time.
This also applies to \name{}, where fast execution is even more challenging due to WebAssembly being a bytecode language and applying our instrumentations at the binary level.

\Cref{table:fuzzing_results} lists the average program executions per second during fuzzing in the column \enquote{Execs/sec}.
With an average speed of 321 executions per second, \name{} is able to quickly explore many paths.
As already described in \Cref{sec:rq-fuzzing-effectiveness}, comparisons between \name{} and AFL are possible in broad strokes only.
This is especially true for performance, because even uninstrumented WebAssembly binaries can execute on average up to 50\% slower than native code~\cite{DBLP:journals/usenix-login/JangdaPBG19}.
Despite this, on benchmark sets one and two, \name{} achieves 414 executions per second on average, which is only 37\% slower compared with 663 native executions per second in AFL.
We believe this is fast enough for practical fuzzing of WebAssembly binaries and respectable, given execution of the target program in a VM.
Finally, as improvements to Wasmtime are orthogonal to our approach, further optimizations of the young VM might also speed up \name{} in the future.

Besides the program execution in a VM, other sources of slowdown in \name{} can come from the applied binary instrumentation.
To evaluate the runtime overhead of the added code, we run the benchmark programs with the test inputs used for RQ3, and compare the runtime of the original, uninstrumented binaries against the runtime when the binaries were instrumented.
The results are shown in the right part of \Cref{table:instrumentation_results}.
On average over 25 program executions, the coverage instrumentation imposes a runtime overhead of 1.46x over the uninstrumented binary.
We will detail the overhead of the canaries in \Cref{sec:rq-efficient-canaries}.
The overhead of the coverage instrumentation is generally higher than for the canaries, because for every branch in the program it adds 13 instructions (\Cref{fig:afl_shim}).
More efficient implementations, e.g., by predicting some branches based on static analysis~\cite{BenKhadra2020}, could further reduce the overhead, which we leave for future work.
The last column of Table~\ref{table:instrumentation_results} shows the combined overhead of both the canary instrumentation and the coverage instrumentation.

\begin{result}
  \name{} performs hundreds of program executions per second, which is only 37\% slower than native AFL, despite executing the program in a VM.
  The coverage instrumentation imposes an average overhead of 1.46x, which dominates the overall overhead imposed by \name{}'s  instrumentation.
\end{result}
\vspace{-1mm}

\subsection{RQ4: Effectiveness of the Canaries in Preventing Exploitation}
\label{sec:rq-effectiveness-canaries}

In the previous research questions, we have analyzed \name{} as an end-to-end WebAssembly fuzzer.
The canary instrumentations from \Cref{sec:canaries} are, however, also useful in a stand-alone setting, namely for catching memory errors in production binaries to prevent exploitation.
To evaluate this scenario, we apply our canary instrumentation to three previously published, vulnerable WebAssembly applications with proof-of-concept exploits~\cite{DBLP:conf/uss/0002KP20}.
The applications use WebAssembly in three different settings: on a website in the browser, on Node.js, and a command-line application for standalone WASI VMs.
Since the canary instrumentation is platform-independent, we can harden binaries in all three settings.
The proof-of-concept inputs exploit two buffer overflow vulnerabilities on the stack, and one buffer overflow on the heap that writes into allocator metadata.
We confirm that the uninstrumented, original WebAssembly binaries can be exploited, which causes cross-site scripting, executes code, and writes to an unintended file, respectively.
Then, we successfully instrument all three binaries, without requiring access to the source code or their build process.
When given correct and benign inputs, those three instrumented binaries work as before, but when passing the exploit inputs, all three examples are successfully terminated by the inserted canary checks.

\begin{result}
  The stack and heap canaries inserted by our binary-only instrumentation effectively hardens existing binaries and protects against previously demonstrated exploits.
\end{result}
\vspace{-1mm}

\subsection{RQ5: Efficiency of the Inserted Canaries}
\label{sec:rq-efficient-canaries}

As demonstrated in the previous section, the inserted canaries can mitigate buffer overflow attacks when applied to existing binaries.
For this usage scenario, it is essential that the canaries have only a minimal impact on performance.
We evaluate their efficiency by running the benchmark programs with and without instrumentation on the inputs from RQ3.
The results are in Table~\ref{table:instrumentation_results}, which shows the execution times with different combinations of canaries relative to the execution time of the uninstrumented programs.

Both the stack and heap canary instrumentation only slightly impact performance, with an average execution time of 1.06x and 1.05x relative to the uninstrumented binary.
The stack canary overhead is similar to efficient implementations of canaries for native binaries~\cite{DBLP:conf/ccs/DangMW15}, which are employed by default in common compilers (Clang, GCC, MSVC).
Some applications, e.g., \textsl{jbig2dec} with an execution time of 1.22x, are impacted significantly more than others, e.g., \textsl{flac}, where the overhead is negligible.
The relative cost of heap canaries depends on the number of memory allocations, especially small ones.
While \textsl{pdfresurrect}, an analyzer of PDF files, stands out due its frequent allocations,
the overhead for the other applications is low or even too small too measure.
The overhead with both canary instrumentations applied (column ``All Canaries'') is approximately the combined overhead of the individual ones, imposing a moderate overhead of 1.11x.

\begin{result}
The overhead imposed by the canary-based oracles is small (1.06x and 1.05x, respectively) and comparable to canary implementations for other languages, which is encouraging for their use to harden production binaries.
\end{result}
\vspace{-1mm}

\section{Related Work}

\parag{Fuzzing}
Out of the many approaches for greybox fuzzing~\cite{manes2019art, DBLP:conf/ccs/KleesRCW018, DBLP:journals/corr/abs-2009-01120, DBLP:journals/corr/abs-1905-10499, DBLP:conf/sp/ChenC18}, we build on the popular \afl{} fuzzer.
Unlike its commonly used GCC and LLVM modes, and the majority of other fuzzers~\cite{rawat2017vuzzer, li2017steelix, peng2018t, aschermann2019redqueen, DBLP:conf/uss/Yun0XJK18}, we do not require source code access, and hence, also cannot rely on compiler-added oracles~\cite{parmesan, DBLP:conf/usenix/SerebryanyBPV12}.
Improvements to the input generation algorithm of \afl{}~\cite{DBLP:conf/woot/MaierEFH20} will also benefit \name{}.

There are several ways to fuzz native binaries.
AFL's QEMU and DynInst modes rely on dynamic instrumentation, which incurs substantial runtime overhead.
Dinesh et al.~\cite{Dinesh2020} propose static instrumentation of x86-64 binaries for fuzzing, but they cannot handle WebAssembly binaries due to the different architecture and also make several assumptions that do not apply in our setting, e.g., position-independent code, which does not exist in WebAssembly, or the presence of relocation information, which we do not require.
Other recent work is about binary instrumentation for coverage~\cite{binary-fuzzing-usenix21}, but does not provide an oracle instrumentation similar to ours.

We know of one approach for fuzzing WebAssembly\footnote{\url{https://github.com/jonathanmetzman/wasm-fuzzing-demo}}, which is a port of LibFuzzer\footnote{\url{https://llvm.org/docs/LibFuzzer.html}}.
They require the source code and only support C and C++ code compiled with Emscripten, which then runs in a browser.
Instead, \name{} is the first to fuzz WebAssembly binaries, with support for WASI applications.
Fuzzing has also been used for testing WebAssembly VMs\footnote{\url{https://github.com/wasmerio/wasmer/tree/master/fuzz} and \url{https://github.com/bytecodealliance/wasmtime/tree/main/fuzz}}, which is orthogonal to fuzzing WebAssembly programs.

\parag{Binary rewriting and overflow protection}
Statically instrumenting native binaries is challenging due to the undecidability of disassembly~\cite{DBLP:conf/pkdd/WartellZHKT11}, with challenges like data inlined in code, variable-length instructions, resolution of indirect jumps, and identification of functions~\cite{DBLP:conf/uss/AndriesseCVSB16}.
Many existing tools rely on the symbol table for recovering function boundaries~\cite{DBLP:conf/usenix/SlowinskiSB12, DBLP:conf/ndss/ChenSABG15, Dinesh2020}.
WebAssembly does not suffer from these problems, a situation we use to our benefit in \name{}.

Several papers have presented binary rewriting techniques for protecting the stack in x86 programs~\cite{DBLP:conf/ndss/ChenSABG15, DBLP:conf/usenix/SlowinskiSB12, DBLP:conf/dimva/NikiforakisPJ13, Dinesh2020}.
BodyArmor inserts instrumentation that monitors reads and writes relative to pointers~\cite{DBLP:conf/usenix/SlowinskiSB12}.
Another technique combines a randomized layout, isolation, and secure allocation for hardening binaries against stack-based vulnerabilities~\cite{DBLP:conf/ndss/ChenSABG15}.
RetroWrite~\cite{Dinesh2020} uses an overflow detection mechanism similar to AddressSanitizer~\cite{DBLP:conf/usenix/SerebryanyBPV12}, i.e., using shadow memory to mark bytes where access is illegal.
These previous three approaches all rely on the symbol table being available.

Prasad et al.\ use a shadow stack for finding buffer overflows reaching across stack frame boundaries, without requiring access to the symbol table~\cite{DBLP:conf/usenix/PrasadC03}.
Their technique is similar to the \name{} canaries in terms of the granularity of detected bugs.
Using a shadow stack does not affect relative references on the stack, and is thus preferable to canaries on x86 where relative references appear.
They suffer from potential false positives and false negatives as most other x86 techniques.

There has also been work on techniques for protecting binaries against heap overflows~\cite{DBLP:conf/lisa/RobertsonKMV03, DBLP:conf/dimva/NikiforakisPJ13, Dinesh2020}.
Robertson et al.~\cite{DBLP:conf/lisa/RobertsonKMV03} and Nikiforakis et al.~\cite{DBLP:conf/dimva/NikiforakisPJ13} both present techniques that, like the \name{} heap canaries, instrument the allocation and deallocation functions such that they insert canaries.
Nikiforakis et al.\ check the canaries at system calls, which means it is likely that overflows are detected early, but at the cost of having to check the canaries often.
We instead opted for the more efficient option of checking canaries during deallocation, to enable efficient fuzzing and low-overhead hardening.

\parag{WebAssembly}
Several papers examine the security of WebAssembly applications and found that exploits, which are no longer possible in native binaries, may still affect WebAssembly~\cite{DBLP:conf/uss/0002KP20, chasm}.
WebAssembly was initially used heavily for malicious cryptomining~\cite{DBLP:conf/ccs/KonothVMLKBV18, DBLP:conf/IEEEares/MuschWJR19, DBLP:conf/imc/RuthZWH18}, but it has recently been shown that today's WebAssembly binaries are rarely malicious~\cite{wasm-empirical}.
While WebAssembly is designed to achieve near native speed, it has been shown to still be around 50\% slower~\cite{jangda2019not}, which explains the performance difference between \name{} and \afl{}.
The Wasabi framework allows for creating dynamic analyses for WebAssembly easily~\cite{asplos2019}, but \name{} uses its own instrumentation for efficiency.

\section{Conclusion}
WebAssembly programs are becoming more and more prevalent, which increases the need for techniques that can uncover security problems.
This paper presents \name{}, the first binary-only greybox fuzzer for WebAssembly.
The approach combines canary-based binary instrumentation to detect overflows and underflows on the stack and the heap, an efficient coverage instrumentation, a WebAssembly VM running the program, and the input generation algorithm of native AFL.
Unlike most other efficient fuzzers, \name{} works directly on production binaries, without requiring access to the source code.
We show that \name{} finds a substantial amount crashes in real-world WebAssembly binaries, while being efficient enough to perform hundreds of executions per second, even though WebAssembly is a slower, non-native language.
Besides as oracles for fuzzing, the canaries also serve as a stand-alone binary hardening technique to prevent exploitation of vulnerable binaries in production.
In this scenario, the approach prevents previously published exploits while imposing low overhead.
Overall, our work is an important step toward securing the increasingly popular WebAssembly platform against exploitation of memory-related vulnerabilities.

\section*{Acknowledgments}
This work was supported by the European Research Council (ERC, grant agreement 851895), and by the German Research Foundation within the ConcSys and Perf4JS projects.

\bibliographystyle{plain}
\bibliography{refs,refsMP}

\end{document}